\documentclass{aa} 

\usepackage{graphicx}
\usepackage{txfonts}
\usepackage[colorlinks=true,citecolor=blue]{hyperref}
\newcommand{\kms}{km s$^{-1}$}
\newcommand{\Msunyr}{M_\odot \, \textup{yr}^{-1}}
\usepackage{multirow}
\usepackage{bm}
\usepackage{natbib}
\usepackage{color}
\usepackage[normalem]{ulem}

\begin{document}

\title{Nonradial and nonpolytropic  astrophysical outflows}
\subtitle{XI. Simulations of the circumstellar environment of RY Tau}

\authorrunning{C. Sauty et al.}

%\email{Raquel.Albuquerque@astro.up.pt christophe.sauty@obspm.fr}
\titlerunning{XI. Simulations of the environment of RY Tau}

\author{
 C. Sauty\inst{1,2}, 
R. M. G. de Albuquerque \inst{1,3,4}, 
V. Cayatte\inst{1}
J. J. G. Lima\inst{3,4}, 
J. F. Gameiro\inst{3,4}}
 
 \offprints{{\tt christophe.sauty@obspm.fr}}
\institute{
Laboratoire Univers et Th\'eories, Observatoire de Paris, Université PSL, Universit\'e Paris Cit\'e, CNRS, F-92190 Meudon, France
\and
Laboratoire Univers et Particules de Montpellier, Universit\'e de Montpellier/CNRS, place E. Bataillon, cc072, 34095 Montpellier, France
\and
Instituto de Astrof\'isica e Ci\^encias do Espa\c{c}o, Universidade do Porto, CAUP, Rua das Estrelas, PT4150-762 Porto, Portugal, 
\and
Departamento de F\'isica e Astronomia, Faculdade de Ci\^encias, Universidade do Porto, Rua do Campo Alegre 687, PT4169-007 Porto, Portugal, 
}

\date{Received 17 September 2021 / accepted 13 June 2022}

\abstract
{Recent observational evidence has shown that RY Tau may present two different  outflow stages, a quiescent and a more active stage. We try to model this phenomenon. }
{We have performed new 2.5D magnetohydrodynamical simulations of the possible accretion-outflow environment of RY Tau based on analytical solutions with the aim to reduce the relaxation time.}
{We used the analytical self-similar solution that we used to model the RY Tau microjet as initial conditions. In the closed field line region of the magnetosphere, we reversed the direction of the flow and increased the accretion rate by increasing the density and velocity. 
We also implemented the heating rate and adjusted it according to the velocity of the flow. The accretion disk was treated as a boundary condition. }
{The simulations show that the stellar jet and the accreting magnetosphere attain a steady state in only a few stellar rotations. This confirms the robustness and stability of self-similar solutions. Additionally, two types of behavior were observed that are similar to the behavior observed in RY Tau. Either the steady stellar outflow and magnetospheric inflow are separated by a low static force-free region or the interaction between the  stellar jet and the magnetospheric accretion creates episodic coronal mass ejections that originate from the disk and bounce back onto the star. }
{The ratio of mass-loss rate to mass-accretion rate that coincides with the change in behavior observed in RY Tau lies within the range of ratios that have been measured during the period in which the initial microjet was analyzed.}


\maketitle

%*************************************************************************
\section{Introduction} \label{sec:intro}
%*************************************************************************

Classical T Tauri stars (CTTS) are pre-main-sequence objects that experience an active evolutionary stage with ongoing accretion and outflow processes. A large fraction of CTTS have jets or outflows, and more than 60 \% have stellar winds \citep{Kwanetal07}. Jets are collimated outflows that may originate from the star, from the surrounding accretion disk, or from the region connecting the disk to the magnetosphere of the star. The current relative consensus that has emerged suggests that the outflow probably has multicomponent sources. The real question is which proportion of each ingredient is relevant and how it may evolve with time. 

One example that of interest for this study is the microjet of RY Tau, which was first observed by \citet{gomezetal01,gomezetal07}. The authors suspected the presence of a small-scale pure stellar jet because the UV emission lines originate in a region that is too small to be produced by a disk wind. \citet{gomezetal07} concluded from the UV density profile that the jet cannot be produced by a magneto\-centrifugally driven disk wind, but is rather of stellar origin. Further observations \citep{StongeBastien08,agra-amboageetal09} confirmed the presence of a micro jet in this faint object. A more recent study involving a spectral and photometric monitoring of RY Tau \citep{babinaetal16} showed evidence of a disk wind perturbed by sporadic magnetospheric ejections. According to \citet{calvetetal04}, RY Tau is a CTTS, whose spectrum indicates that it is a G1. It has a stellar radius of $2.9\pm 0.4 R_\odot$ and a stellar mass of $2.0 \pm 0.3 M_\odot$. In this study, we focus on the inner microjet and ask whether most of this jet comes from the star or from the magnetosphere.

Many models have been developed in order to explain the angular momentum loss in CTTSs and how accretion is linked to outflows. For instance, \citet{MattPudritz05, MattPudritz08} suggested that accretion processes feed stellar winds and have an important role in angular momentum extraction, while the disk-locking mechanism seems to fail, as shown by \cite{Romanovaetal11}.

This has been confirmed in \citet{sautyetal11}, where an analytical model of the microjet succeeded to reproduce the observations and the total mass loss rate extracted from the polar coronal hole of the star. Using observational measurements of the radius, mass, and period of the star, of the mass-loss rate, the terminal velocity, and the radius of the jet, we have constrained the solution parameters using the self-similar approach of \citet{sautyTsinganos94}. 

Much numerical effort has been expended to model outflowing structures that cause the star to spin down, in particular, on inner disk winds originating at the boundary of the  disk and the magnetosphere. This is the original idea of X-winds \citep{shuetal94}, although fan-shaped winds are unlikely to exist \citep{BardouHeyvaerts96}. The effort was extended to conical winds (\citet{Romanovaetal09}, to 3D simulations in \citet{Romanovaetal11}, or to a Rex-Wind \citep{ZanniFerreira13}. The structure studied by \citet{ZanniFerreira13} suggests that magnetospheric ejections, similar to the coronal mass ejections observed on the Sun, are efficient enough to extract a relevant amount of angular momentum. The advantage of these simulations is that they take the accretion disk structure and the disk wind into account in a consistent way, together with the magnetospheric accretion and ejection. 
However, all simulations fail to take the inner stellar jet component into account because they neglect any source of heating in the inner stellar corona.

\cite{Orlandoetal11} reported a similar magnetospheric mass ejection triggered by flares. However, they considered much lower mass-accretion rates of a few $10^{-9}\Msunyr$ that fit better Class III objects, such as weak-line T Tauri Stars (WTTS).

More recent 3D simulations of the interaction of the disk and the magnetosphere with disk winds and magnetospheric ejections have been performed  \citep{Pantolmosetal2017,Pantolmosetal2020,Irelandetal2021}, showing that the disk-locking spins up the star. This spin-up can be compensated only by a strong wind either from the star or the magnetosphere, or from both locations.

Instead of using {\it \textup{ad hoc}} magnetospheric configurations and waiting for the relaxation time for the system to attain equilibrium, we take advantage of the existence of an analytical solution for the stellar jet of RY Tau and use this solution as the initial condition for our simulations, both for the stellar jet and for the magnetospheric accretion. This technique of using analytical solutions relies on a method that  has been developed in the past to study the propagation of multicomponent jets, for instance, by \cite{Matsakosetal2008,Matsakosetal2009,Matsakosetal2012}.

In the magnetospheric accretion zone, we play a simple trick using the reversibility of the velocity in steady MHD equations. Thus we reverse the direction of the flow and increase the density and the velocity such that we can play on the accretion rate for a given initial mass loss. As the magnetosphere is magnetically dominated, the density and velocity of the flow are less important in the geometry than the magnetic field. As mentioned above, other simulations have been performed by other groups such as \citet{Romanovaetal09} and \citet{ZanniFerreira13}, but they emphasized the connection between the disk wind and the magnetospheric ejections. We instead focus on the effect of the stellar jet on magnetospheric ejections. 

One main drawback of our approach so far is that we do not model the inside of the accretion disk with a fully consistent disk wind. We are currently working on these improvements, but the description of the interior of the disk requires implementing resistivity. Nonetheless, in order to highlight some of the basic features of the interaction of the stellar jet with the magnetospheric wind, it is sufficient to treat the disk as a boundary condition. The disk is the equatorial source of mass and magnetic fluxes. Thus, we perform here new simulations that include the stellar component, the magnetospheric accretion flux tubes, and the central equatorial static dead zone in the magnetospheric atmosphere of the star. The accretion disk is treated as an equatorial boundary condition.

In Sec. \ref{sec:technic} we explain the simulation setup and how we built on analytical solutions. Then we present two general classes of solutions showing that we have mainly two types of behavior of the magnetospheric wind in Sec.\ref{sec:results}. We also discuss that these two classes of solutions are independent of the simulation conditions, except for the ratio of mass-accretion rate to mass-loss rate. We compare our results in Sec. \ref{sec:discussion} to other simulations and to observations of RY Tau.  In Sec. \ref{sec:conclusion} we conclude with the application of our results to the dichotomic behavior of RY Tau.

%-----------------------------------------------------------
\section{Numerical conditions of the  simulations}\label{sec:technic}
%-----------------------------------------------------------
%-----------------------------------------------------------
\subsection{Analytical solution that was used as initial setup}\label{subsec:initialsetup}
%-----------------------------------------------------------

We performed simulations with the PLUTO code developed by Andrea Mignone and collaborators (see \cite{mignoneetal07}) in its version 4.2, with a Riemann solver following a Lax-Friedrichs scheme (TVDLF). We satisfied the Courant-Friedrichs-Levy (CFL) condition by using a CFL number of 0.4. We used self-similar solutions developed by \cite{sautyTsinganos94} to build the initial setup and the boundary conditions, as explained below.

\citet{sautyetal11} modeled the RY Tau microjet using a semianalytical solution derived from the meridional self-similar approach of  \citet{sautyTsinganos94}.  These analytical solutions were developed for jets from low-mass accreting T Tauri stars (TTS). They are exact solutions of the steady ideal MHD equations for the mass density, $\rho$, the pressure $P$, the velocity field $\vec{V,}$ and the magnetic field $\vec{B}$ of one fluid. These equations describe the motion of the protons in a highly ionized plasma.

The authors took the stellar parameters of RY Tau into account, namely the mass-loss rate, the stellar radius, and the mass: $\dot{M}_{loss}=10^{-8.5}\Msunyr$ \citep{gomezetal01}, $R_\star=2.4 R_\odot$ , and $M_\ast=1.63 M_\odot$ \citep{hartiganetal95}, respectively.  \citet{calvetetal04} measured slightly different values with a stellar mass between  $1.7$ and $2.3 M_\odot$. This does not change our results, which can easily be scaled with mass. The velocity scales as the square root of the mass, which means that varying the mass from $1.63$ to $2.3 M_\odot$ increases the velocity, the mass flux, and the magnetic field strength by 19\% for a constant density. Similarly, the observed mass-loss rate that was used to fix the boundary mass density on the star can be rescaled to be adapted to model other asymptotic mass-loss rates. We took advantage of the stability of the first analytical solution given in \citet{sautyetal11} to use it as initial conditions for the simulations. From the observational constraints, we deduced an analytical solution with a typical stellar dipolar magnetic field of about 600 G, which is reasonable for a T Tauri star. The UV electron density observed at various distances fits the obtained solution within the error bars. 

The analytical solution also gives as an output an equatorial rotational velocity of $8.6$ \kms, or a period of roughly $14.2$ days \citep{sautyetal11}.
This value remains within the lower limit of measured velocities for RY Tau in \cite{bouvieretal93}. However, this may be rather low for actual values of $v\sin i$ around $52$ \kms, which  correspond to a period about 3 days when the inclination of the system of $66^o$ is taken into account; see \cite{petrovetal99}. However, \cite{Petrovetal21} found a time variability in H$\alpha$ and NaID2 lines with a period of about 22 days, which is not related to the rotational period of the star. Nevertheless, the obtained solution reproduces the observational constraints on the micro jet seen in UV spectral lines derived by \citet{gomezetal01,gomezetal07}  fairly well, and those derived  in the optical by \citet{StongeBastien08} and \citet{agra-amboageetal09}. The dynamics of the outflow is not affected strongly by the rotation of the star because the stellar jet is pressure driven.

\begin{figure}[tbp]
\includegraphics[width=\columnwidth]{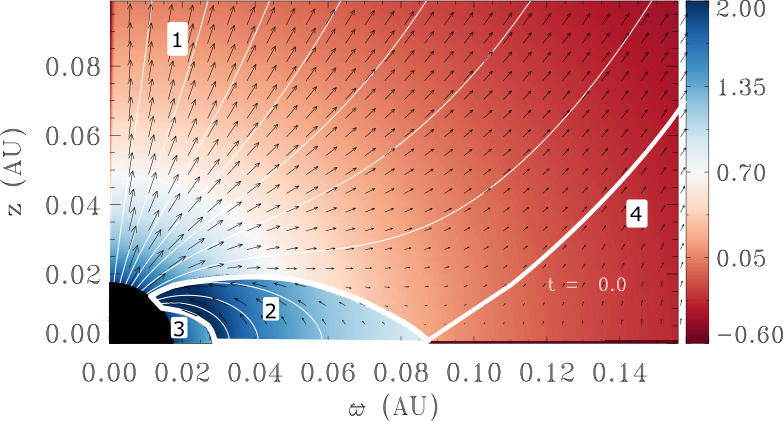}
\caption{Illustration of the initial conditions. The color scale shows the
logarithm of the density, the arrows show the velocity, and the solid white lines represent
the magnetic field. From the polar axis (vertical axis) to the equatorial
plane (horizontal axis), three different initial regions lie above the stellar
photosphere and one region lies outside of the magnetosphere. The first region is an outflow from the star that emerges from the polar coronal hole. It is surrounded by the second region, in which the flow velocity has been reversed and the density enhanced to induce magnetospheric accretion. The second region starts on the equatorial plane from $0.03$ AU to $0.09$ AU. Inside this region lies a third region, corresponding to a dead or static zone without flow. It is located above the equatorial plane between the radius of the star at $0.011$ AU and the inner disk radius $0.03$ AU. The fourth region is the disk wind. It starts on the right of the last open field-line connected to the magnetosphere at $\varpi(\pi/2) = 0.09$ AU on the equatorial plane (solid white line) and extends above the accretion disk.} 
\label{Fig.t0}
\end{figure}

Thus, the initial conditions rely on the RY Tau micro jet model presented in \citet{sautyetal11}. The initial conditions of our simulations are given in Fig. \ref{Fig.t0}. There are four regions, namely the stellar jet component, the magnetospheric accretion region, the dead zone, and the disk wind.  Formally, the meridional self-similar model is only valid for the star-driven jet (region 1). 
In the closed field-line region (region 2), the flow given by the analytical solution starts from the star and assumes a sink of material at the equatorial boundary. This solution exhibits a deceleration of the outflow in this region. Because the MHD equations are reversible, we can change the direction of the flow in the solution and still satisfy these equations. By doing so, we obtain an inflow in the closed field-line region that is accelerated from the equatorial plane to the star and simulates the magnetospheric accretion flow.

The accretion region corresponds to the actual accretion columns. However, the system being axisymmetric, it has the shape of a helmet streamer directed toward the star. A first set of simulations was conducted allowing the disk to reach the star. We ran a second set of simulations that reproduced accretion disks with a truncation radius.  In this set, we created a static dead zone (region 3) in which the magnetic field of the star was strong enough to limit the inner radius of the disk.

Additionally, the analytical solution provides a surrounding disk wind emerging from an infinitely thin accretion disk (region 4). Although this wind may be weak and have a negligible impact on the dynamics, it is an important feature of the simulation. The disk wind rotation profile is not Keplerian in the analytical solution. However, we performed two types of simulations, one in which we kept the original analytical rotation profile, and a second in which we imposed a Keplerian profile for the rotation of the disk and the disk wind. We discuss below that this Keplerian profile has qualitatively no influence on the dynamics of the solutions. 

%--------------------------------------------------------
%           FIGURE t0 FIG 1
%--------------------------------------------------------
%--------------------------------------------------------

The most complete simulations included the four different initial regions we just discussed. In region 1 of Fig. \ref{Fig.t0}, the outflow from the star emerges from the polar coronal hole. The stellar jet component is entirely given by the analytical solution, including heating, without any change. 

Region 2 of Fig. \ref{Fig.t0} extends along the equatorial plane from $0.03$ AU to $0.09$ AU. It coincides with the magnetospheric accretion zone, in which the density is higher. In this region, we imposed a reversal of the sign of the poloidal velocity, $\bm{V}_{\rm p}$, of the analytical solution to induce accretion, as mentioned above. We have an accelerated transonic inflow from the disk to the star, which remains subAlfv\'enic everywhere in the closed fieldline region, however. This means that we can start close from an equilibrium state, even in the initial closed field-line region.  Furthermore, the azimuthal component of the magnetic field must also reverse for consistency with the isorotation law. This reversal of the azimuthal magnetic field is required because we do not wish to reverse the azimuthal velocity on the star. The directions of stellar and the disk rotation should be consistent. In region 2, we can uniformly multiply the velocity by a given fixed factor. The geometry of the flow is dominated by magnetic forces. It is not affected by changes in the dynamics. The same is true for the mass density, which we can uniformly multiply by a chosen factor. Thus, we can control and increase the initial magnetospheric accretion rate, consistently with observations. 

In summary, we multiplied the analytical density and velocity in the closed field-line accreting region by some given factors (negative for the velocity), and we reversed the toroidal magnetic field while keeping the analytical values of the pressure, the poloidal magnetic field, and the toroidal velocity. These initial values are also the boundary conditions imposed on the equatorial plane where the flow connects to the disk.

The static region or dead zone (region 3) starts along the equatorial plane, from above the stellar surface at $0.01$ AU up to the inner disk radius at $0.03$ AU. This static dead zone has nothing to do with the so-called dead zone in the inner part of the accretion disk at large distances. This is more like the dead zone inside helmet streamers in the solar corona.  In this dead zone, the poloidal velocity of the flow is zero in the initial state. The fluid is in solid rotation with the constant rotation frequency of the star at a given latitude. To be consistent with the isorotation law, the azimuthal component of the magnetic field must also be set to zero in this region, because, being static, there is no angular momentum transport inside it.

Region 4 is the disk wind. It starts at the right end of the magnetosphere at $0.09$ AU along the equatorial plane.  In this region, we either initially set the rotational velocity to the analytical value, or we replaced it by a Keplerian profile in the entire region, depending on the simulation. This basically amounts to saying that a disk wind emerges from the infinitely thin equatorial disk, but its density remains weak compared to the density in the accreting magnetosphere and compared to the stellar jet density.

%---------------------------------------------------------
\subsection{Relaxation method and mapping the heating}\label{subsec:evol}
%---------------------------------------------------------

The analytical solution corresponds to a given map of the heat rate deposit along the flow. In the comoving frame and in steady state, this rate is given by 
\begin{equation}
q_{An}= \rho_{\rm An} \vec{V}_{\rm p,An} \cdot \left[\nabla \left({\Gamma \over \Gamma -1} {P_{\rm An} \over \rho_{\rm An}}\right)
- {{\nabla P_{\rm An}} \over {\rho_{\rm An}}} 
\right]\,,
        \label{eq:theq_AN}
\end{equation}
where the subscript $_{An}$ refers to the values of the analytical solution. The total heating rate, $q$, corresponds to the heating rate, $H$, minus the cooling rate, $\Lambda$. $P$ is the pressure, $\rho$ is the density, and $\vec{V}_{p}$ is the poloidal velocity. $\Gamma$ is the adiabatic index, which is $5/3$ for a monoatomic gas. 

As the equation of energy is written in the comoving frame, changes in the direction and magnitude of the poloidal velocity compared to the analytical solution should be taken into account in the heating rate. Thus, the total heating rate $q$ in the simulation box was calculated from the following expression:
\begin{equation}
        q = q_{\rm An}\frac{\vec{V}_{\rm p} \boldsymbol{\cdot} \vec{V}_{\rm p,An}}{V_{\rm p,An}^2} \, .
        \label{eq:theq_new}
\end{equation}

The time-dependent energy equation is given by
\begin{equation}
\frac{\partial P}{\partial t}
+\vec{V}\cdot{\nabla P}+\Gamma P\nabla\cdot{\vec{V}} =q(\Gamma-1)  \, .
        \label{eq:theq}
\end{equation}

 In order to solve the energy equation, PLUTO uses a fractional step formalism with an operator splitting. The partial derivative of the pressure is split into two terms,
     \begin{equation}
\frac{\partial P}{\partial t} = {\left.{\partial P \over \partial t}\right|_{\rm heat}} + {\left.{\partial P \over \partial t}\right|_{\rm MHD}}  \,  .
        \label{eq:splitting}
\end{equation}

The first term in the above equation is given by
\begin{equation}
\left.{\partial P \over \partial t}\right|_{\rm heat} 
%=\Lambda(t) 
= q(\Gamma-1)
\,.
\label{eq:heat}
\end{equation}

Then the PLUTO code calculates the second term using the MHD module 
without heating (see PLUTO userguide.pdf, sections 6.2 and 9.1), 
\begin{equation}
\left.{\partial P \over \partial t}\right|_{\rm MHD} = - \vec{V}\cdot{\nabla P} - \Gamma P\nabla\cdot{\vec{V}} 
\,.
\label{eq:MHD}
\end{equation}
 Adding Eq. (\ref{eq:heat}) to Eq (\ref{eq:MHD}), we solve Eq. (\ref{eq:theq}) and obtain the total pressure.

The method is a relaxation method for a given map of the heating, which is just adjusted to the velocity pattern, considering a quasi-equilibrium. Thus, it is not surprising that in a few stellar rotations only, the simulation reaches its final state, as discussed below.

%---------------------------------------------------------
\subsection{Boundary conditions}\label{subsec:evol}
%---------------------------------------------------------

PLUTO was used in its static grid version and in spherical coordinates. Along the polar axis, we used axisymmetric conditions, which correspond to the symmetry conditions there (no reflection, no perpendicular gradients, etc.). 

 The Alfv\'en surface of the analytical solution is a sphere of radius $r_{\rm Alf}=22.4 r_\odot=0.104$ AU \citep{sautyetal11}. The radius of the star is $r_\star=0.011$ AU.
 We started the simulations at $r_o$ just above $r_\star$.
 There is a small zone between the polar axis and the last open fieldline and from $r_o$ up to a radius $r_1$, below two stellar radii, in which gradients are very sharp and difficult to calculate.
 In this zone, physical quantities were kept constant equal to their initial values in order to avoid too strong gradient calculations, which are time consuming. From now on, this zone is called the "fixed zone".
 We explored various sizes of this zone (e.g., Tab. \ref{tab:PLUTO_simulation_config}). We followed the same procedure as described in \cite{BogovalovTsinganos05} and \cite{Graciaetal05}, but in a very restricted area.  If this zone is too small, the computation usually stops because the time step becomes too small. The initial acceleration of the stellar jet is extremely stable around the polar axis and remains almost unchanged when the size of this zone is changed. 

The boundary conditions for the density, the pressure, the poloidal velocity and the toroidal magnetic field were set free, at $r_o$ in the closed fieldline region, between the last open field-line and the equator.
The toroidal velocity and the poloidal magnetic field 
were fixed and equal to their initial values, however.
At the boundaries of the dead zone, we fixed solid-rotation conditions, which implies that the toroidal magnetic field value is zero (see Sec. \ref{subsec:initialsetup}). 

Along the equatorial plane, the boundary conditions were kept fixed and given by the initial setup. The rotation of the disk outside the magnetospheric region (equatorial plane of region 4 in Fig. \ref{Fig.t0}) was either fixed by the analytical solution values or corresponded to a Keplerian velocity.

%---------------------------------------------------------
\subsection{Numerical setup}\label{subsec:numericalsetup}
%---------------------------------------------------------

We assumed a planar symmetry with respect to the equatorial plane, and we used a grid in colatitude $\theta$ of $128$ points between  $\theta=1/128$ rad and $\theta=\pi/2-1/128$ rad. In the simulations, the accretion disk therefore is a thin equatorial layer with an opening angle $\theta_{\rm disk}=1/128$ rad.

First, we performed simulations with two grids (see the last column of Tab. \ref{tab:PLUTO_simulation_config}). The first grid has $512$ points between $r_o=0.016$ AU and $r=0.18$ AU. The second grid is a stretched logarithmic grid of $256$ points between $r=0.18$ AU and $r=25$ AU.

Second, we performed simulations with higher spatial resolution in the radial direction $r$. There, we used three consecutive adjacent grids (see the last column of Tab. \ref{tab:PLUTO_simulation_config}). We still had a uniform  grid of $512$ points, starting at $r=0.02$ AU and finishing $r=0.18$ AU. We kept the stretched grid of $256$ points between $r=0.18$ AU and $r=25$ AU. We introduced an additional uniform grid of $64$ points between $r_0=0.012$ AU and $r=0.02$ AU.

The PLUTO code uses a unique cell in the azimuthal direction, with periodic conditions due to axisymmetry.
We used two different intervals for the fixed zone. In the lowest-resolution simulations, we used $[r_0,r_1]=[0.016,0.020]$ AU. In the two last simulations with higher resolution, we brought the fixed zone closer to the star by using $[r_0,r_1]=[0.012,0.018]$ AU (see columns 6 and 7 of Tab. \ref{tab:PLUTO_simulation_config}).
Changing the number of grids and/or the radius of the base and/or the size of the fixed zone $[r_0,r_1]$ 
modifies the computational time and the relaxation time. 
The higher the resolution, the slower the time step. 

The PLUTO code uses dimensionless quantities. We normalized all quantities by the values given in \cite{sautyetal11}. The normalization factor for the velocity was $V_{\rm PLUTO}=111$ \kms. This is the Alfv\'{e}n velocity, $V_{Alf}$, along the polar axis of the analytical solution scaled to RY Tau. The normalization factor for the radius was $r_{\rm PLUTO}=r_{\rm Alf}$, which implies $t_{\rm PLUTO}=r_{\rm PLUTO}/V_{\rm PLUTO} \simeq 1.7$ days. Similarly, the normalization factor for the mass density was $\rho_{\rm PLUTO}=2.48\times 10^{-15}$ g cm$^{-3}$.

%---------------------------------------------------------------
% TABLE  1  NEW TABLE
%---------------------------------------------------------------
\begin{table*}[htp]
\caption[Configuration of the simulations.]{ Configuration of the selected simulations. The simulations are identified in the first column, followed by the initial multiplying factors imposed for velocity ($V$) and density ($\rho$) with respect to the analytical solution in region 2 of Fig. \ref{Fig.t0}. Some simulations include a dead zone (column 4) and a Keplerian rotating disk (column 5). Columns 6 and 7 define the limits of the fixed zone. Depending on the starting radius, a different number of adjacent grids has been used in the radial direction (last column).}
\label{tab:PLUTO_simulation_config}
\centering
\begin{tabular}{cccccccccc}
\hline
Simulation ID & $V$  & $\rho$ & Dead & Keplerian  & $r_0$ & $r_1$ & Number of grids  \\
 & factor & factor & Zone?  & Rotation? & in AU & in AU & in radial direction  \\
\hline
Case A & $-1.0$ & $1.0$ & No & No & 0.016 & 0.020 & 2 \\
Case B & $-1.5$ & $1.5$ & No & No & 0.016 & 0.020 & 2   \\
Case C & $-1.5$ & $5.0$ & No & No & 0.016 & 0.020 & 2   \\
Case D & $-2.0$ & $10.0$ & No & No  & 0.016 & 0.020 & 2   \\
Case C.1 & $-1.5$ & $5.0$ & Yes & Yes & 0.016 & 0.020 & 2   \\
Case D.1 & $-2.0$ & $10.0$ & Yes & Yes & 0.016 & 0.020 & 2   \\
Case C.2 & $-1.5$ & $5.0$ & Yes & Yes &  0.012 & 0.018 & 3   \\
Case D.2 & $-2.0$ & $10.0$ & Yes & Yes  &  0.012 & 0.018 & 3   \\
\hline
\end{tabular}
\end{table*}

%---------------------------------------------------------------

We performed several simulations to make a parametric study. We present a selection of the most representative
ones in 
Tab. \ref{tab:PLUTO_simulation_config} .

A first series of simulations (cases A, B, C, and D) allowed us to see the effect of increasing the mass-accretion rate. 
These simulations do not include the Keplerian rotation profile on the disk. Thus, we kept the analytical disk wind solution in region 4 of Fig. \ref{Fig.t0}. They started at $r_o=0.016$ AU, and the dead zone was not included. The magnetospheric accretion region filled the closed field-line region, and so the inner part of the disk reached the star through a boundary layer. However, in our axisymmetric simulations, the dead zone is completely embedded in the optically thick magnetospheric accretion zone. Thus, an external observer would not be able to conclude whether there is a dead zone or not.

A second series of simulations (cases C.1, D.1,  C.2, and D.2) included the dead zone and the Keplerian rotation. Cases C.1 and D.1 started at $r_o=0.016$ AU with two radial grids and a fixed zone up to $r_1=0.020$ AU for two different mass-accretion rates. Cases C.2 and D.2 were improved simulations starting closer to the star at $r_o=0.012$ AU and with a reduced fixed zone up to $r_1=0.018$ AU. This implies that we also increased the resolution by using three radial grids. 
In this second series of simulations, the initial truncation radius at which the dead zone (region 3) stops was equal to $0.03$ AU.
 The relaxation state was reached before $30$ PLUTO time units. The static dead zone was conserved after a compression leading to a truncation radius of nearly $0.022$ AU for cases C.1 and C.2, and less than $0.02$ AU for cases D.1 and D.2.

Our simulations were also performed with an increased resolution and larger grid scales compared to previous works of other authors. 
For comparison, \cite{Irelandetal2021} had $256$ points in colatitude between $[0,\pi]$, equivalent to our grid. They only had $320$ points on a logarithmic grid from 2 solar radii ($\sim 0.02$ AU) and $101$ solar radii ($\sim 0.47$ AU), however. \cite{Romanovaetal09} had typical grids of $31$ points in colatitude between $[0,\pi/2]$. The grids in the radial direction had $51$ points between $2$ to $32$ stellar radii for simulations in the conical regime. For their typical CTTS, this is between $\sim 0.019 $ AU and $\sim 0.30$ AU. The grids had $85$ points between $2$ to $96$ stellar radii ($\sim 0.018 $ AU to $\sim 0.89$ AU) for simulations in the propeller regime. \cite{ZanniFerreira13} used a grid of $100$ points in colatitude between $[0,\pi/2]$, and $214$ stretched points in the radial direction between one and $28.6$ stellar radii ($\sim 0.009$ AU to $\sim 0.27$ AU).

%************************************************************
\section{Two classes of solutions} \label{sec:results}
%************************************************************

%--------------------------------------------------------
%           FIGURE Test A,B,C,D
%--------------------------------------------------------
\begin{figure*}
\centering
\begin{tabular}{cc}
    \includegraphics[width=0.48\linewidth]{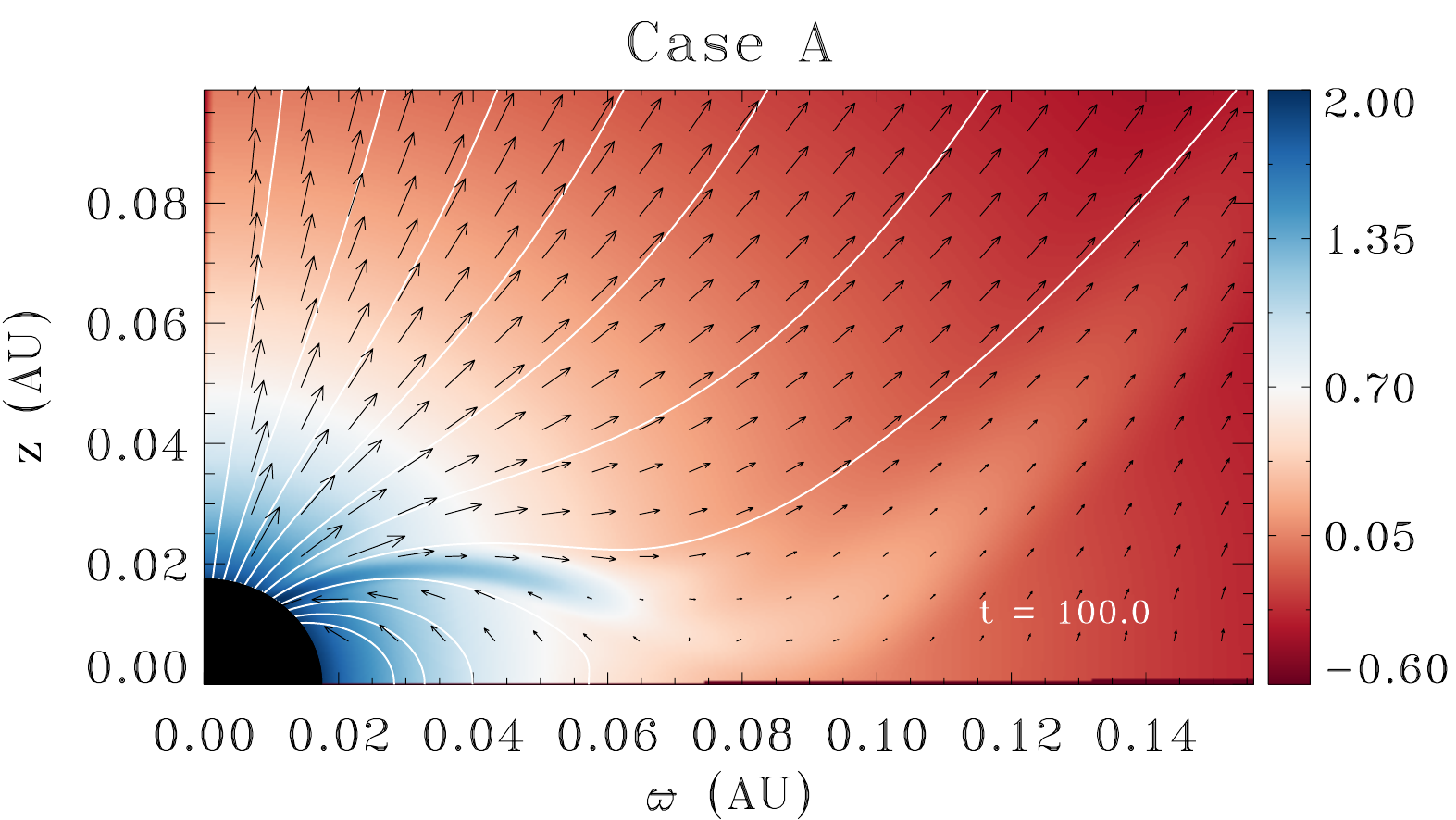} &   \includegraphics[width=0.48\linewidth]{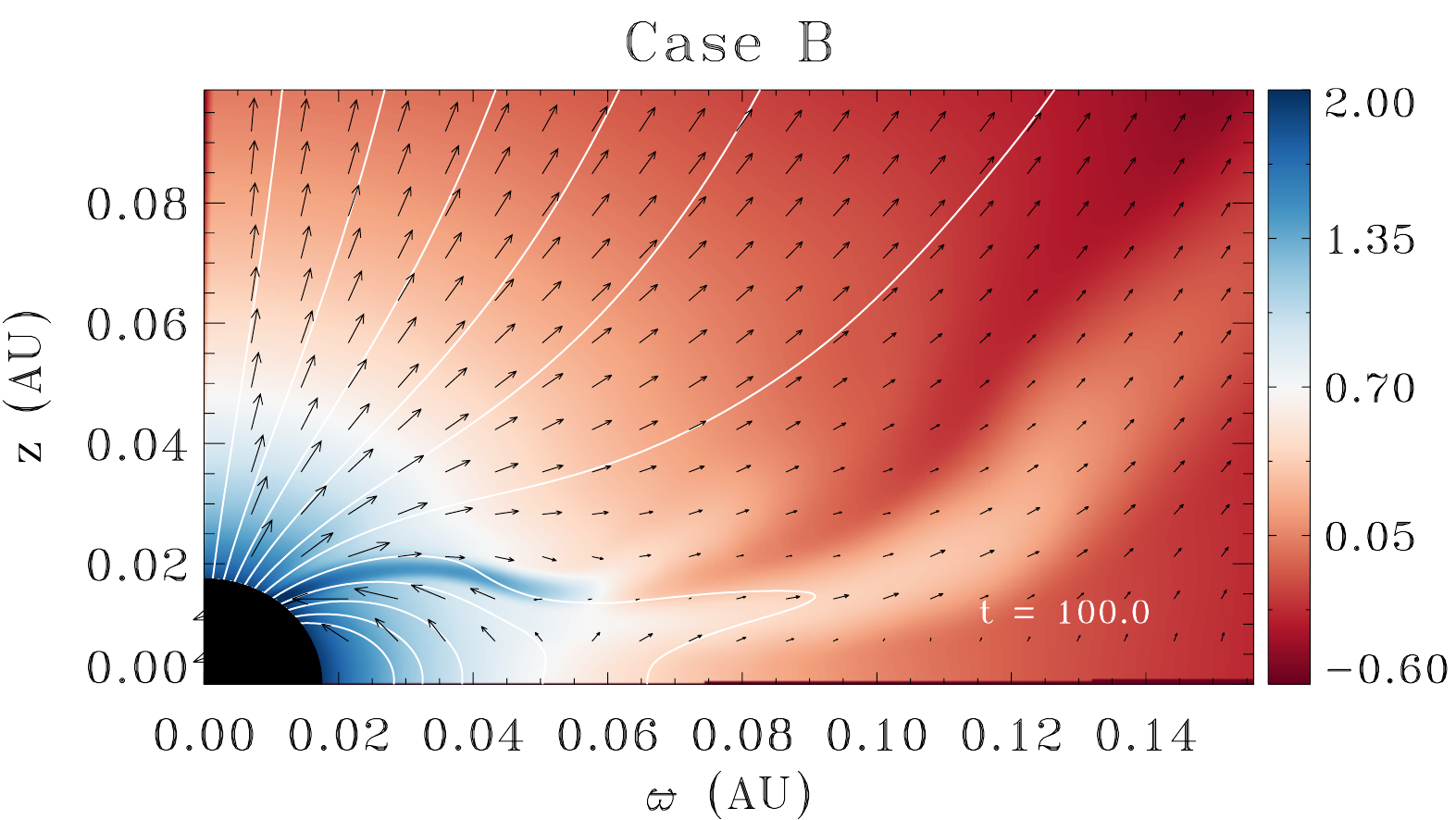} \\
(a) & (b) \\[6pt]
    \includegraphics[width=0.48\linewidth]{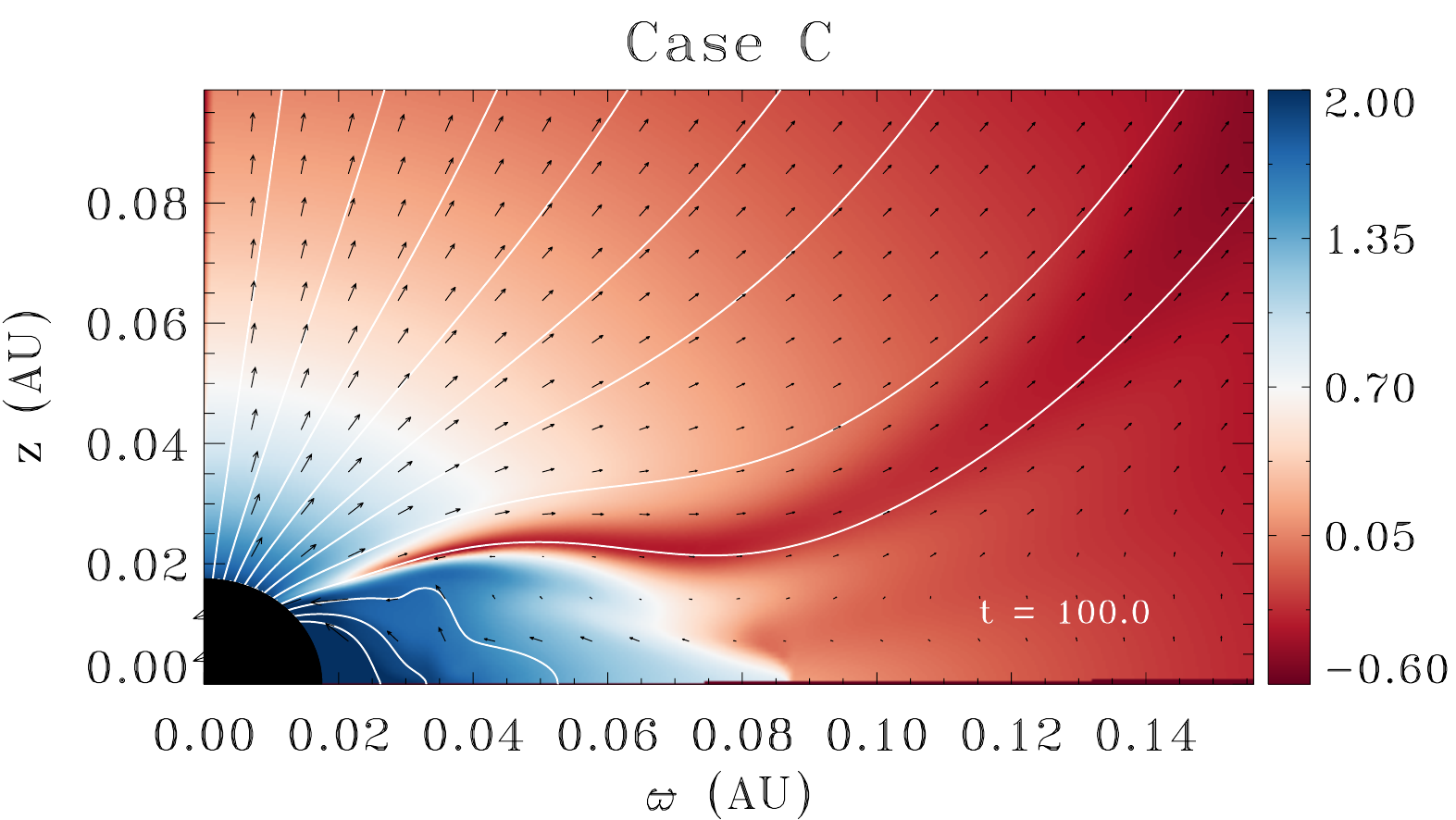} &   \includegraphics[width=0.48\linewidth]{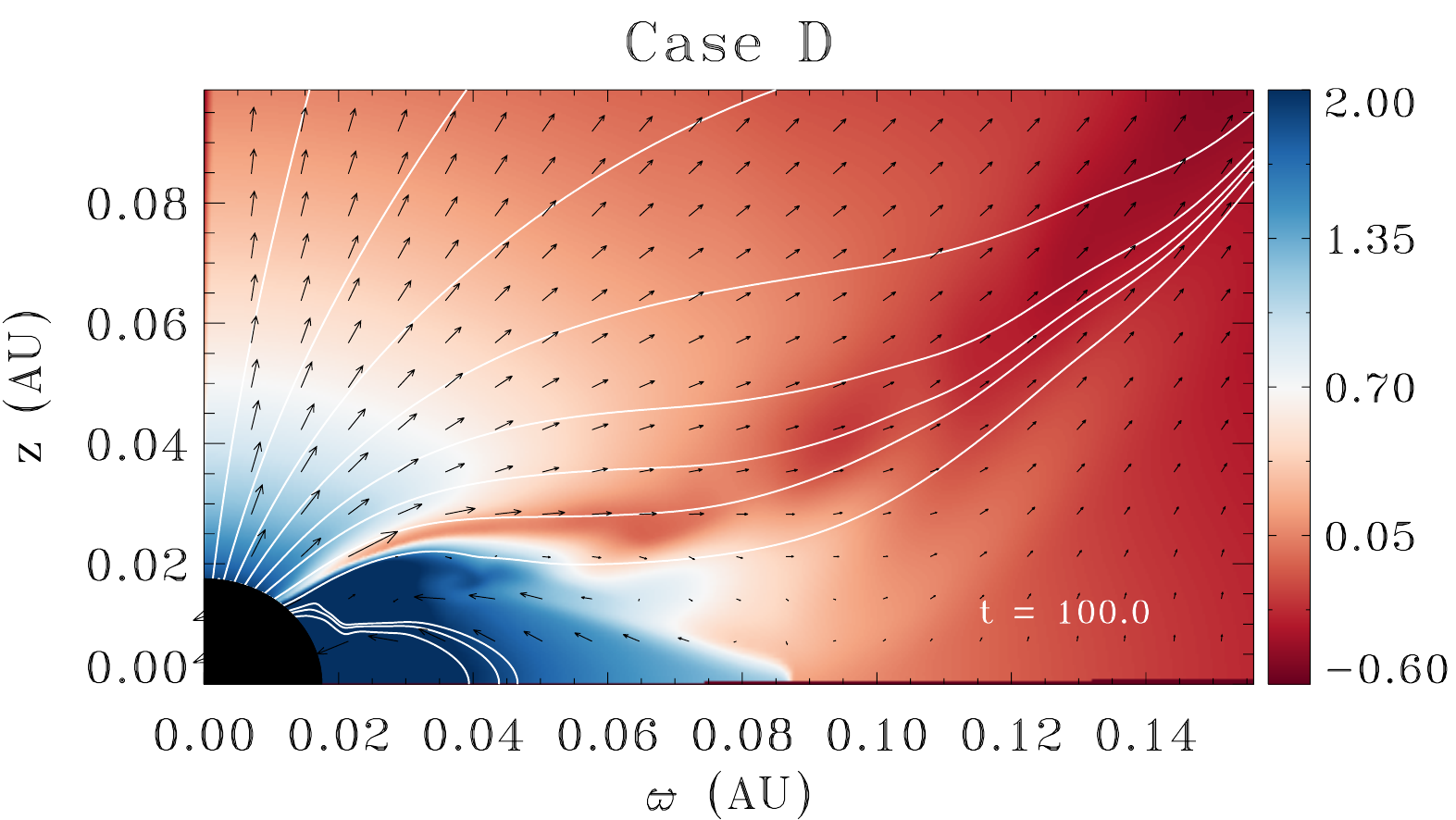} \\
(c) & (d) \\[6pt]
\end{tabular}
\caption{Density maps for simulations A, B, C, and D. The plots represent the logarithmic density in PLUTO units ($\rho_{\rm PLUTO}=2.48\times 10^{-15}$ g cm$^{-3}$). The velocity vectors are shown through the black arrows, and the magnetic field lines are represented by the solid white lines. The distances on the vertical ($z$) and horizontal ($\varpi$) axis are represented in astronomical units. All the plots refer to PLUTO time = 100, which corresponds to approximately 12 stellar rotations.
\label{FigTestABCD}}
\end{figure*}
%--------------------------------------------------------

%--------------------------------------------------------
%           FIGURE Test C.1 time evolution
%--------------------------------------------------------
\begin{figure*}
\centering
\begin{tabular}{cc}    \includegraphics[width=0.48\linewidth]{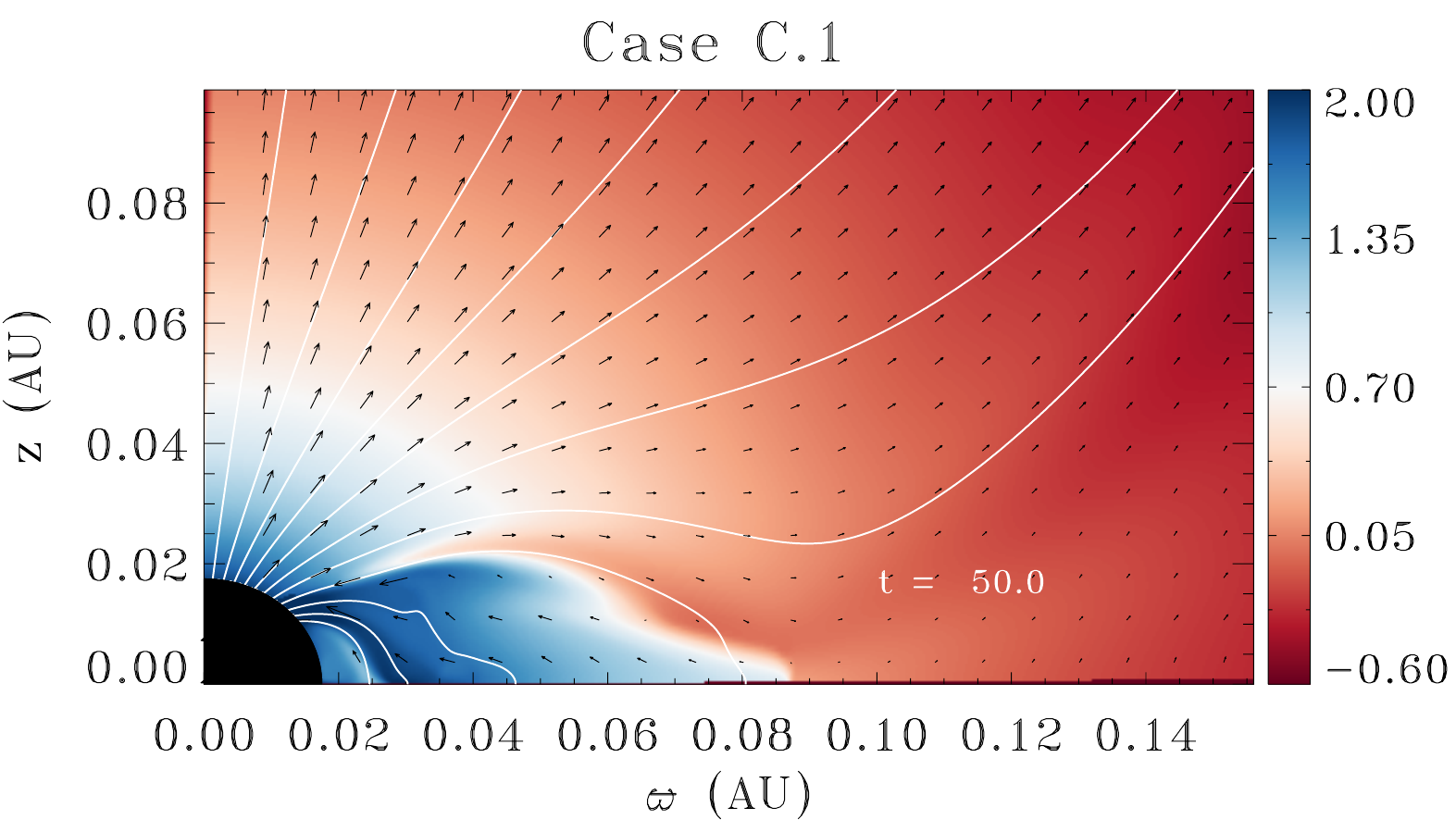} &   \includegraphics[width=0.48\linewidth]{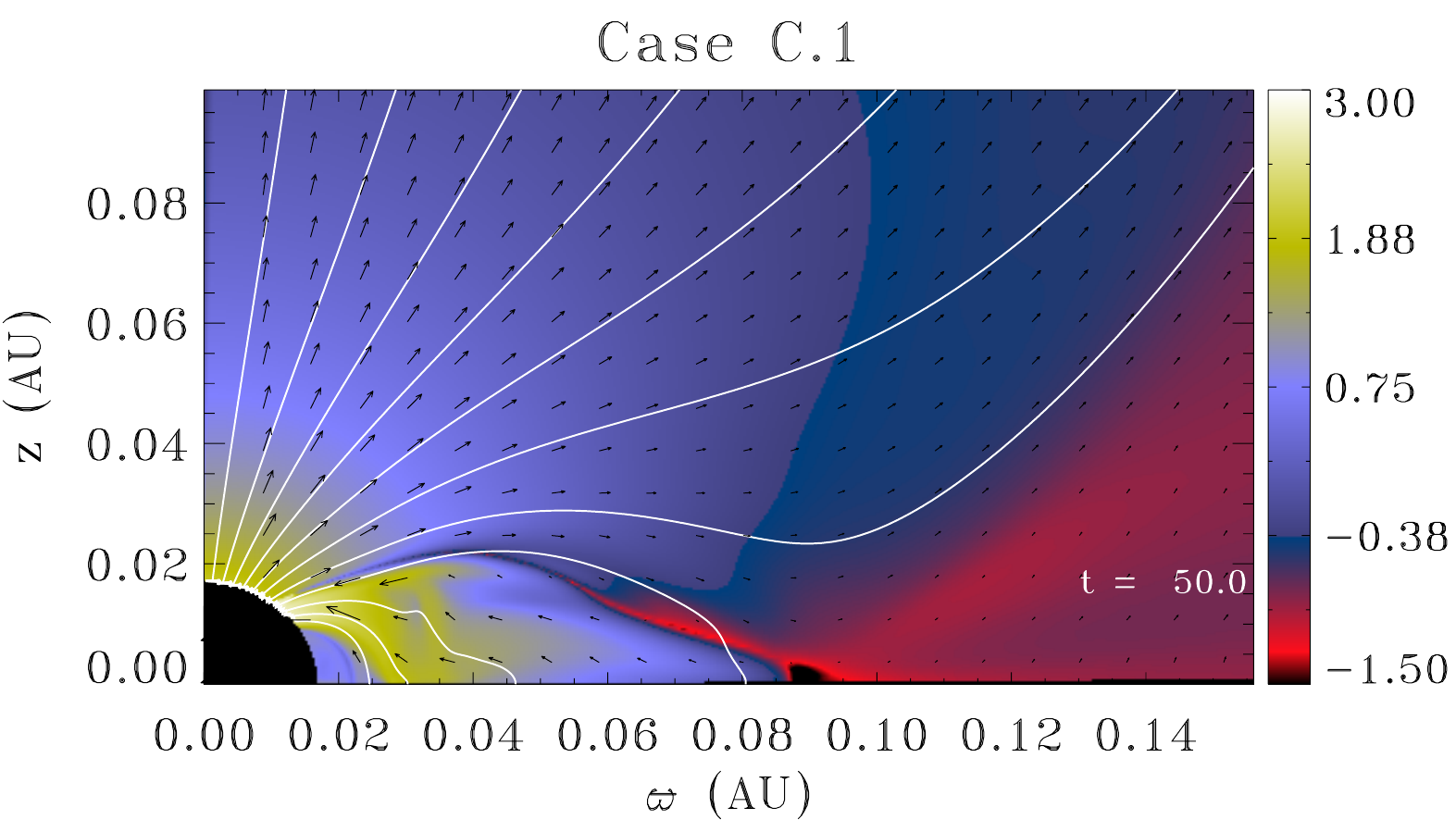} \\
(a) & (b) \\[6pt]
    \includegraphics[width=0.48\linewidth]{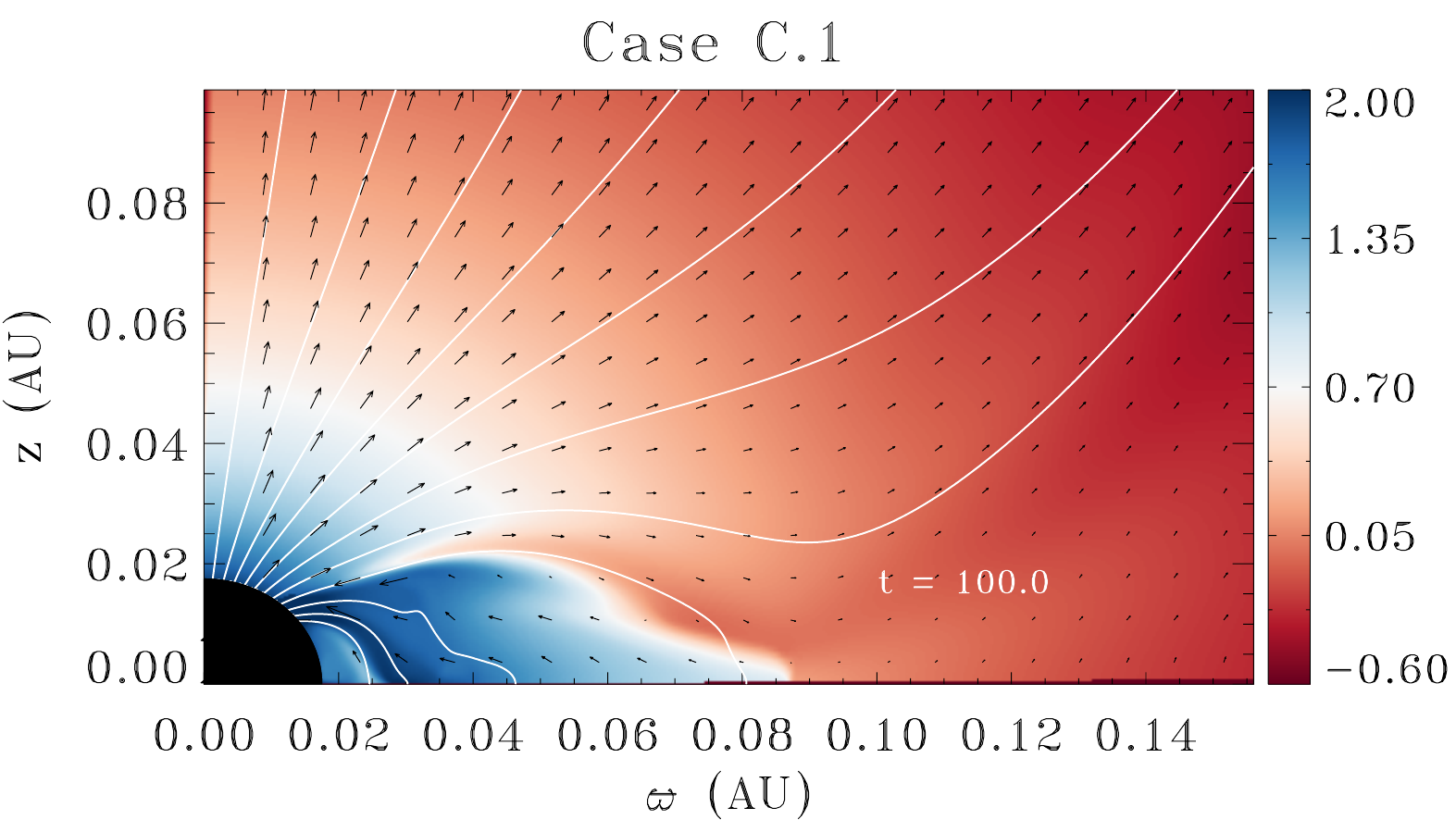} 
&   \includegraphics[width=0.48\linewidth]{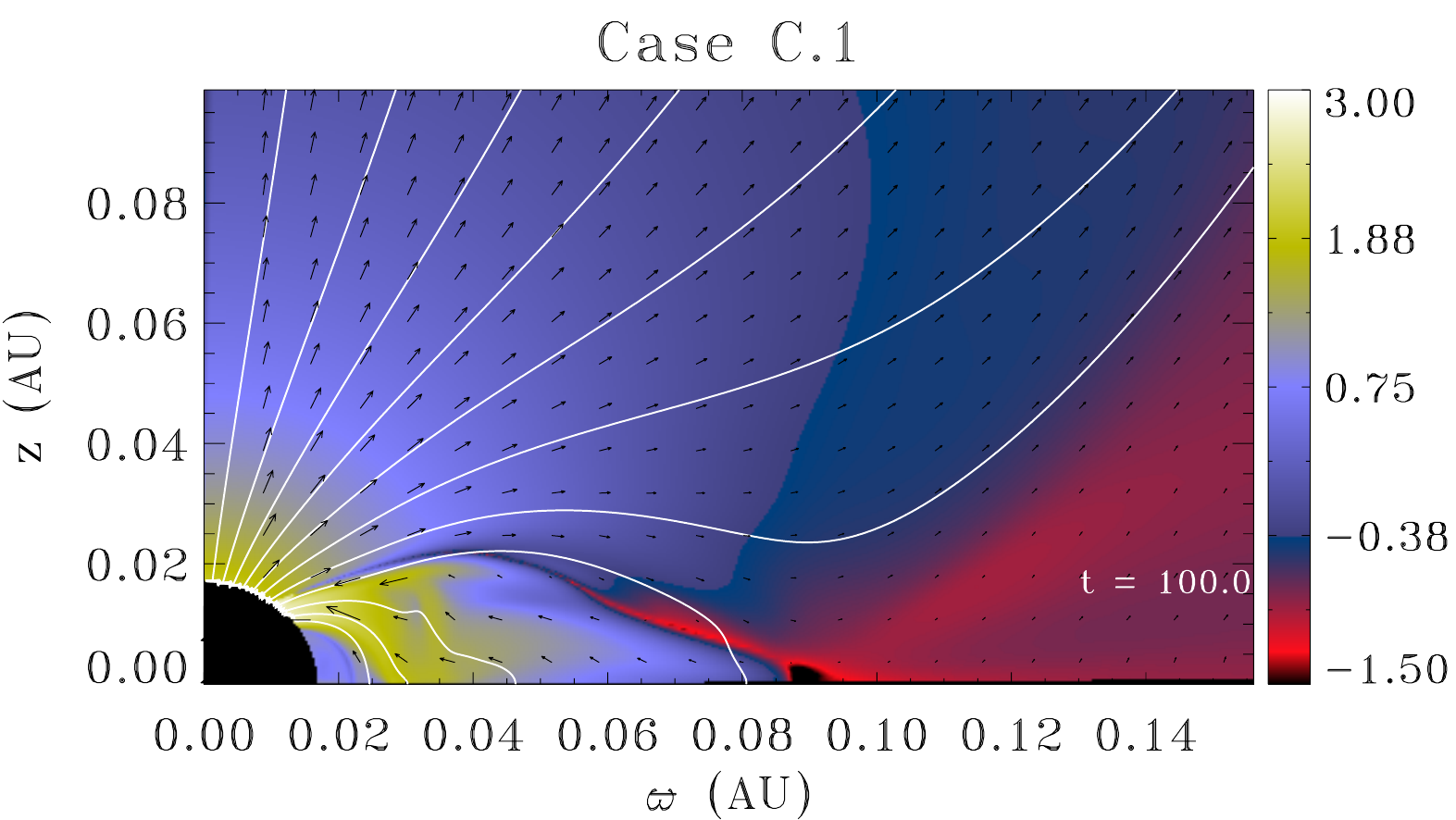} \\
(c) & (d) \\[6pt]
\end{tabular}
\caption{Logarithmic density maps  in PLUTO units ($\rho_{\rm PLUTO}=2.48\times 10^{-15}$ g cm$^{-3}$; left) and logarithmic mass flux maps, also in PLUTO units ($\rho V_{\rm PLUTO}=2.75\times 10^{-8}$  g cm$^{-2}$ s$^{-1}$; right). We plot simulation case C.1 at  PLUTO time = 50 and  100.  Case C.1 includes a dead zone. The numbers are identical at the two different times. The velocity vectors are shown as black arrows, and the magnetic field lines are represented by the solid white lines. The distances on the vertical ($z$) and horizontal ($\varpi$) axis are represented in astronomical units.}
\label{FigtimeevolC.1}
\end{figure*}
%--------------------------------------------------------

%************************************************************
\subsection{Typology of the solutions} \label{sec:introresults}
%************************************************************

%--------------------------------------------------------
%           FIGURE Test D.1 time evolution
%--------------------------------------------------------
\begin{figure*}
\centering
\begin{tabular}{cc}
    \includegraphics[width=0.48\linewidth]{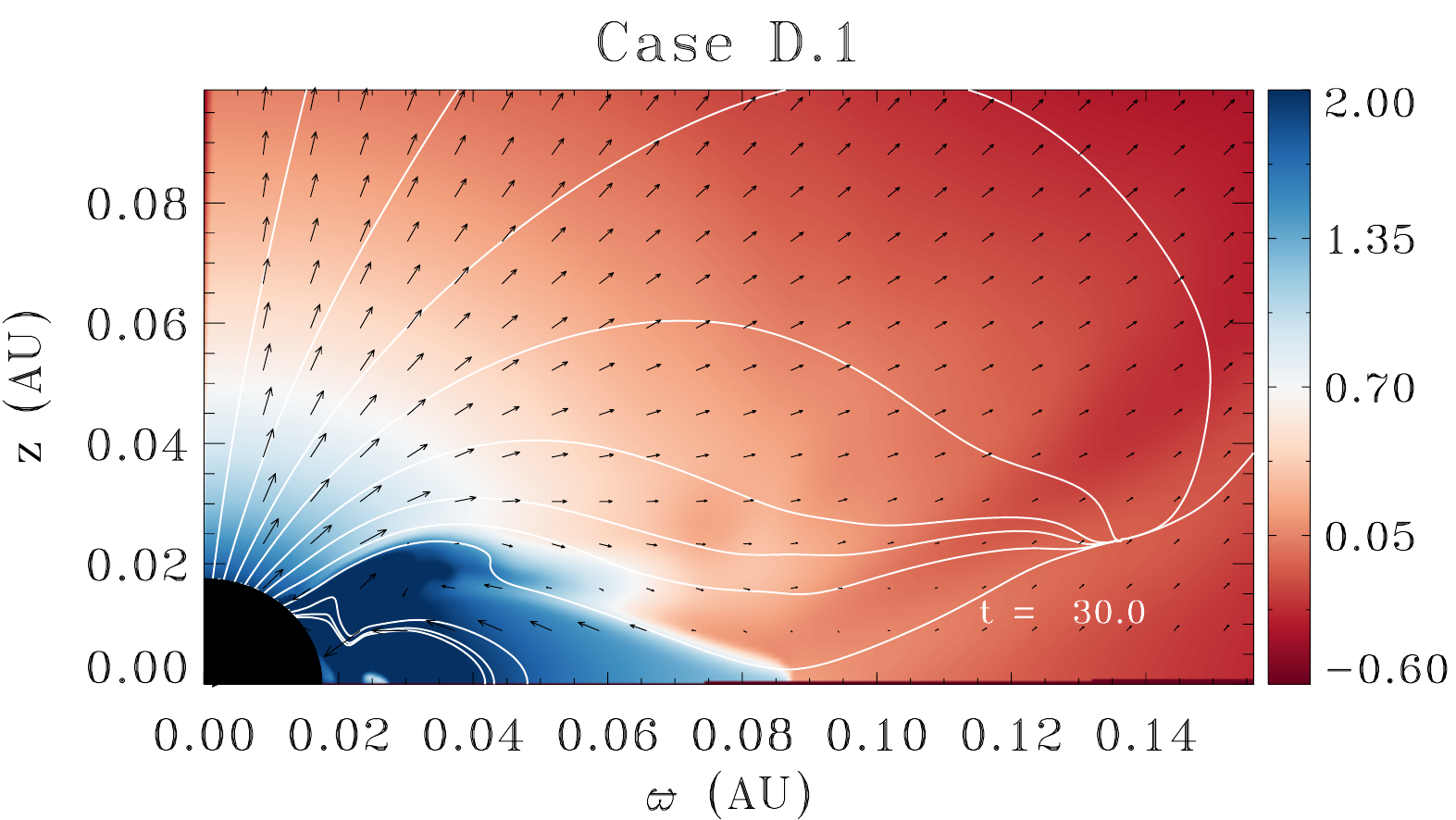} 
&   \includegraphics[width=0.48\linewidth]{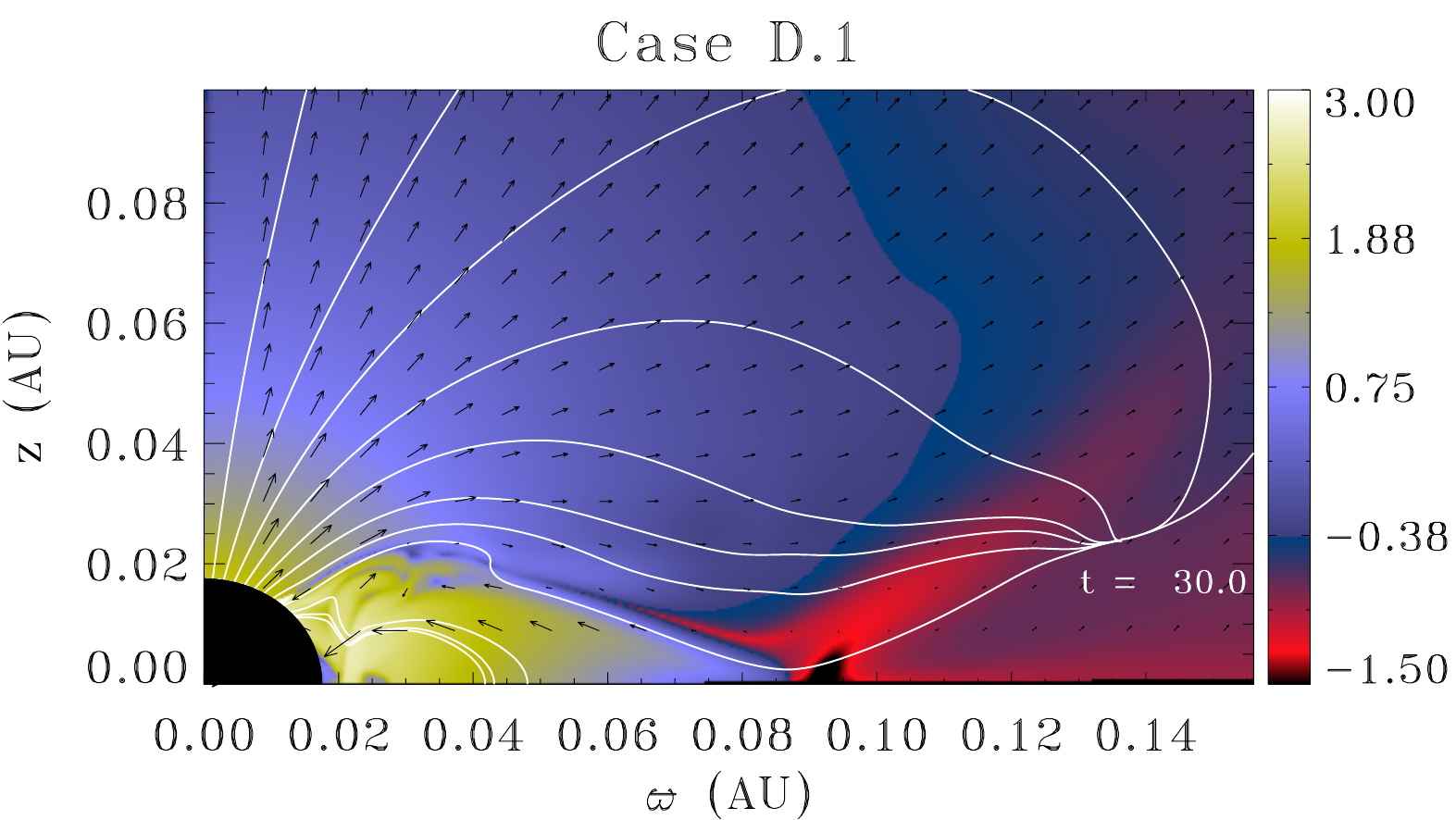} \\
(a) & (b) \\[6pt]
    \includegraphics[width=0.48\linewidth]{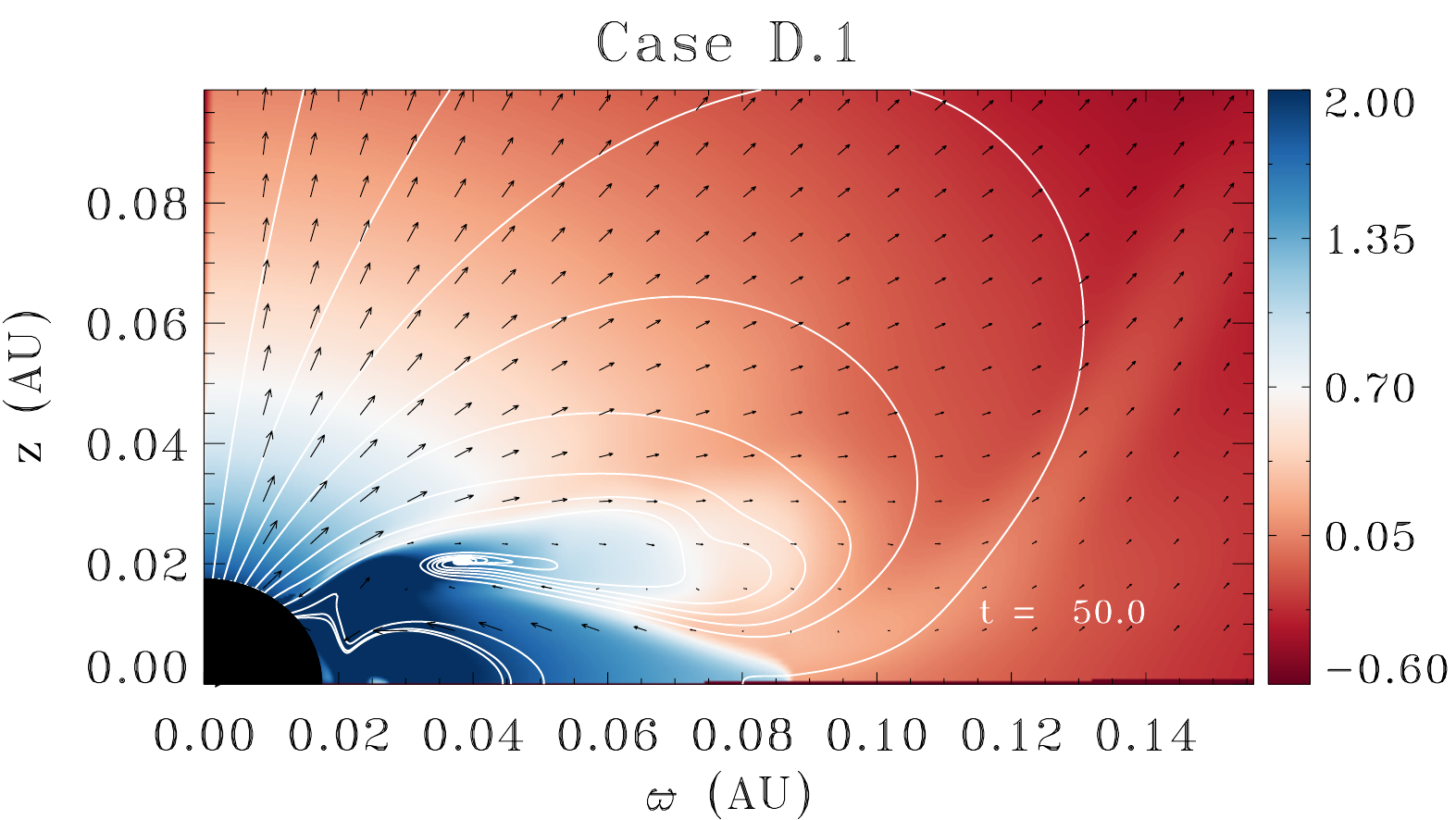} 
&   \includegraphics[width=0.48\linewidth]{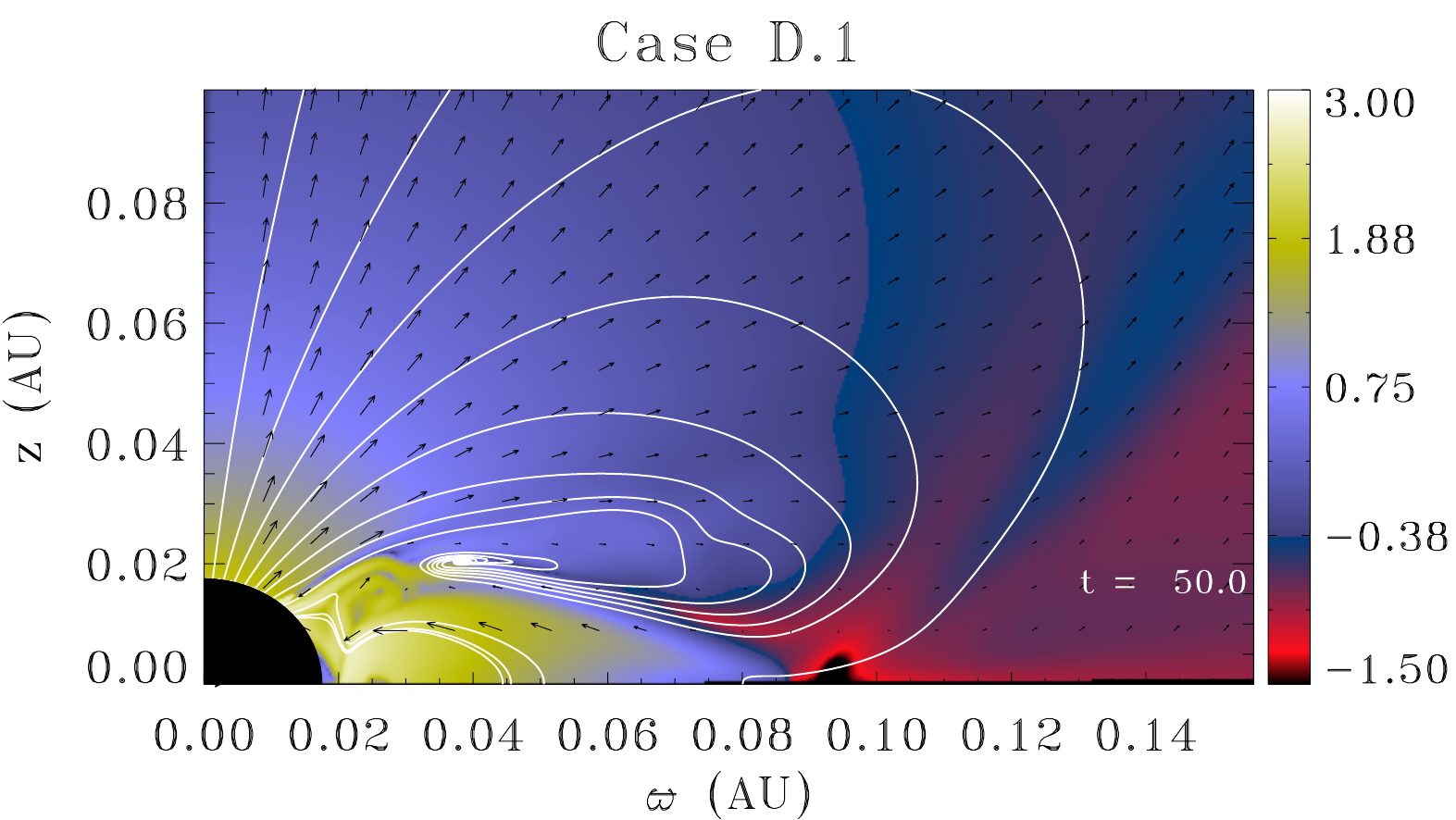} \\
(c) & (d) \\[6pt]
    \includegraphics[width=0.48\linewidth]{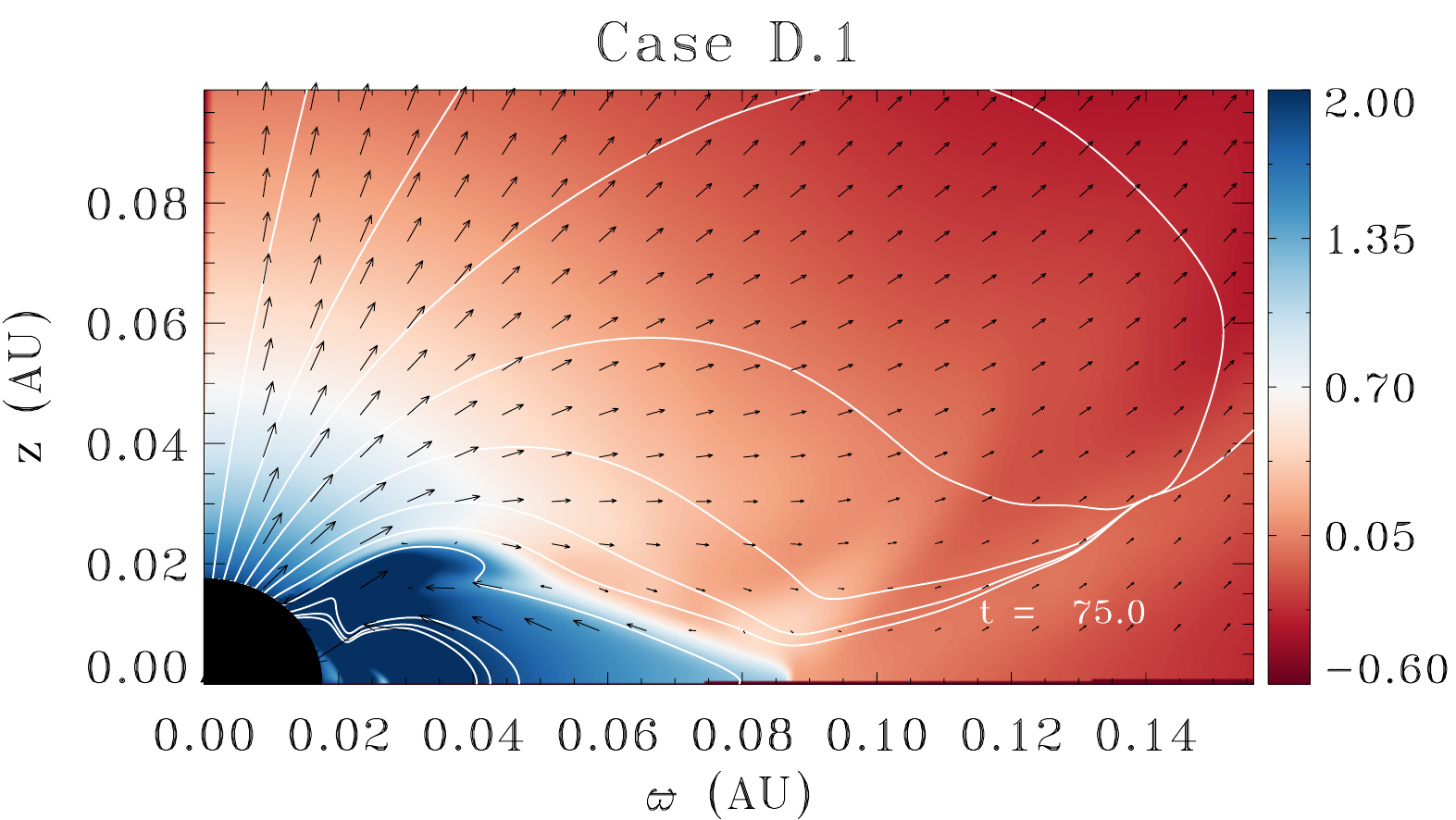} 
&   \includegraphics[width=0.48\linewidth]{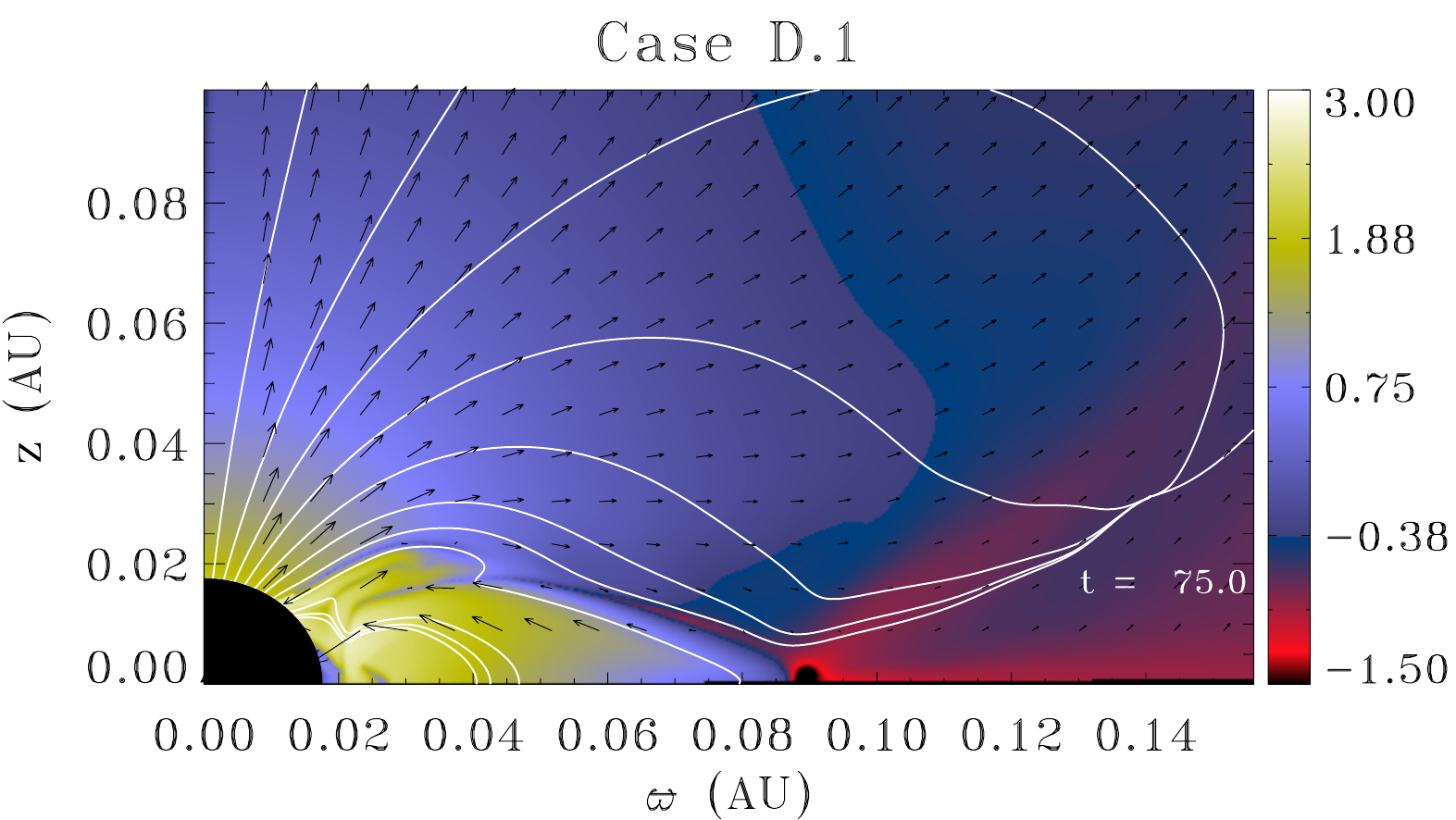} \\
(e) & (f) \\[6pt]
    \includegraphics[width=0.48\linewidth]{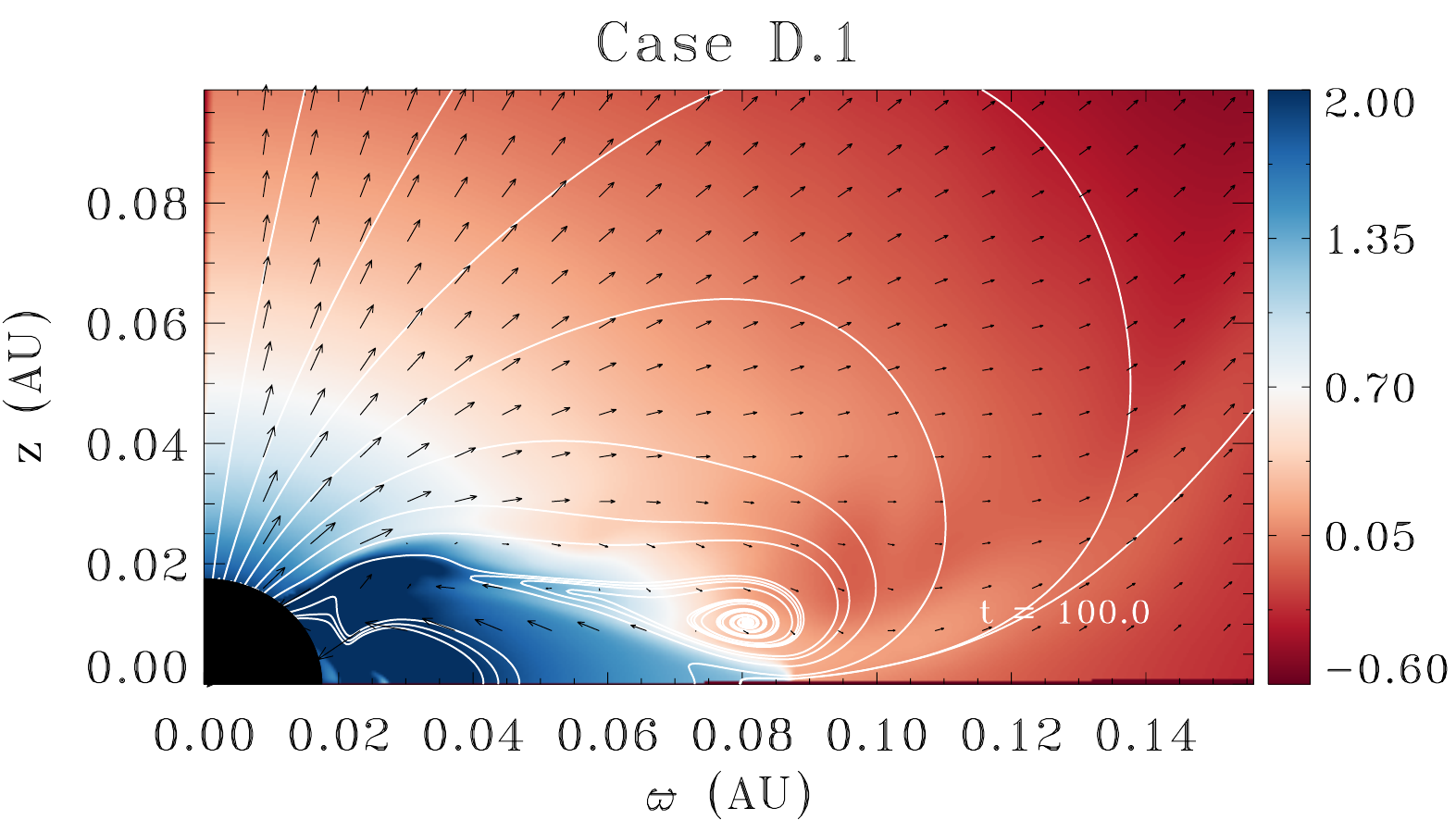} 
&   \includegraphics[width=0.48\linewidth]{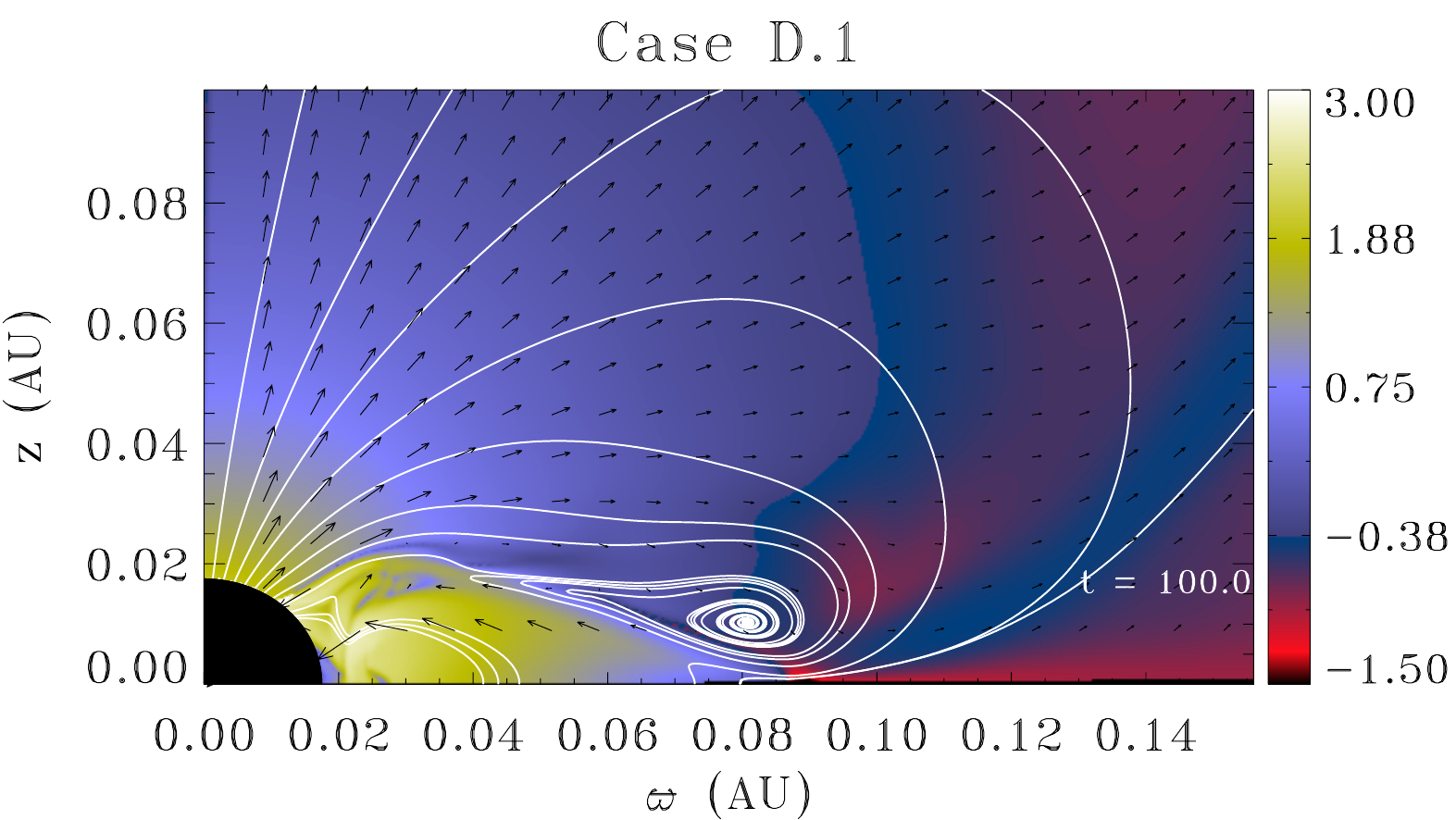} \\
(g) & (h) \\[6pt]
\end{tabular}
\caption{Logarithmic density maps  in PLUTO units ($\rho_{\rm PLUTO}=2.48\times 10^{-15}$ g cm$^{-3}$; left) and logarithmic mass flux maps, also in PLUTO units ($\rho V_{\rm PLUTO}=2.75\times 10^{-8}$  g cm$^{-2}$ s$^{-1}$; right). We plot simulation case D.1 at  PLUTO time = 30, 50, 75, and 100. The velocity vectors are shown as black arrows, and the magnetic field lines are represented by the solid white lines. The distances on the vertical ($z$) and horizontal ($\varpi$) axis are represented in astronomical units.}
\label{FigtimeevolD.1}
\end{figure*}
%--------------------------------------------------------

The different setups we present here show how our self-similar analytical solution can be used to model, in presence of a stellar jet, an accreting magnetosphere with different mass-accretion fluxes by varying the initial accretion velocities and densities. We present the evolution of the solution typologies as a function of the multiplying factors of $\rho$ and $V$ in the accreting magnetosphere.

Velocity and density cannot be increased arbitrarily in the simulations. In other words, there is a correspondence between the combination of multiplying factors for velocity and density. If these two parameters are not increased in a specific way, the simulation does not reach a stable state.

For the series of cases A, B, C, and D, we plot the density maps at PLUTO time $\simeq 100$ in Fig. \ref{FigTestABCD}. In all the solutions, the inner stellar jet around the polar axis is conserved. The mass loss and the extracted angular momentum from the star remain close to the original values of the analytical solution. This confirms the stability of self-similar solutions regardless of the situation around the equatorial plane, as reported in  \cite{Sautyetal17} and \cite{Todorovetal16}.

Our results highlight two types of simulations. We observe a first type with lower mass-accretion rates (cases A, B, and C) that reach steady state without episodic magnetospheric flows. A second type of simulations appears for higher mass-accretion rates (case D), in which the flow reaches a quasi-steady state with episodic magnetospheric mass ejection. Cases A and B show a small plume of enhanced density, however, that stays at the frontier between the stellar jet and the magnetospheric accretion.

For simulations with a dead zone, we found the same transition between cases C.1 (Fig. \ref{FigtimeevolC.1}) and D.1 (Fig. \ref{FigtimeevolD.1}). 
These figures show the evolution of the two types of solutions. The first (case C.1) is completely steady, while the second (case D.1) is quasi-steady.  
Increasing the resolution of the simulations and starting at a lower radius does not change the results. We always end up with two typical classes of solutions represented by C.2 and D.2, with similar characteristics to cases C.1 and D.1, respectively. 
Case C.2 is presented in Fig. \ref{FigC.2} and case D.2 in Fig. \ref{FigD.2}. The two solution types reach their final state after 25 PLUTO time units, which means after three rotations of the central star. 

%************************************************************
\subsection{Steady solutions without magnetospheric ejection} \label{sec:expected}
%************************************************************
This first type of solutions (cases A, B, C, C.1, and C.2) came as a surprise as the solutions do not have magnetospheric ejecta. Previous simulations with a magnetic configuration like this always show such magnetospheric mass ejection, for instance,{\it } in \cite{Romanovaetal09} or \cite{ZanniFerreira13}. Even the propeller regime versus conical wind regime observed in the simulations of \cite{Ustyugovaetal2006} is completely different from our study. There the conical wind of magnetospheric origin is always present. The fast axial jet appears only if the star is a fast rotator.

%--------------------------------------------------------
%           FIGURE C.2 FIG 
%--------------------------------------------------------
\begin{figure*}
\centering
\begin{tabular}{cc}
    \includegraphics[width=0.48\linewidth]{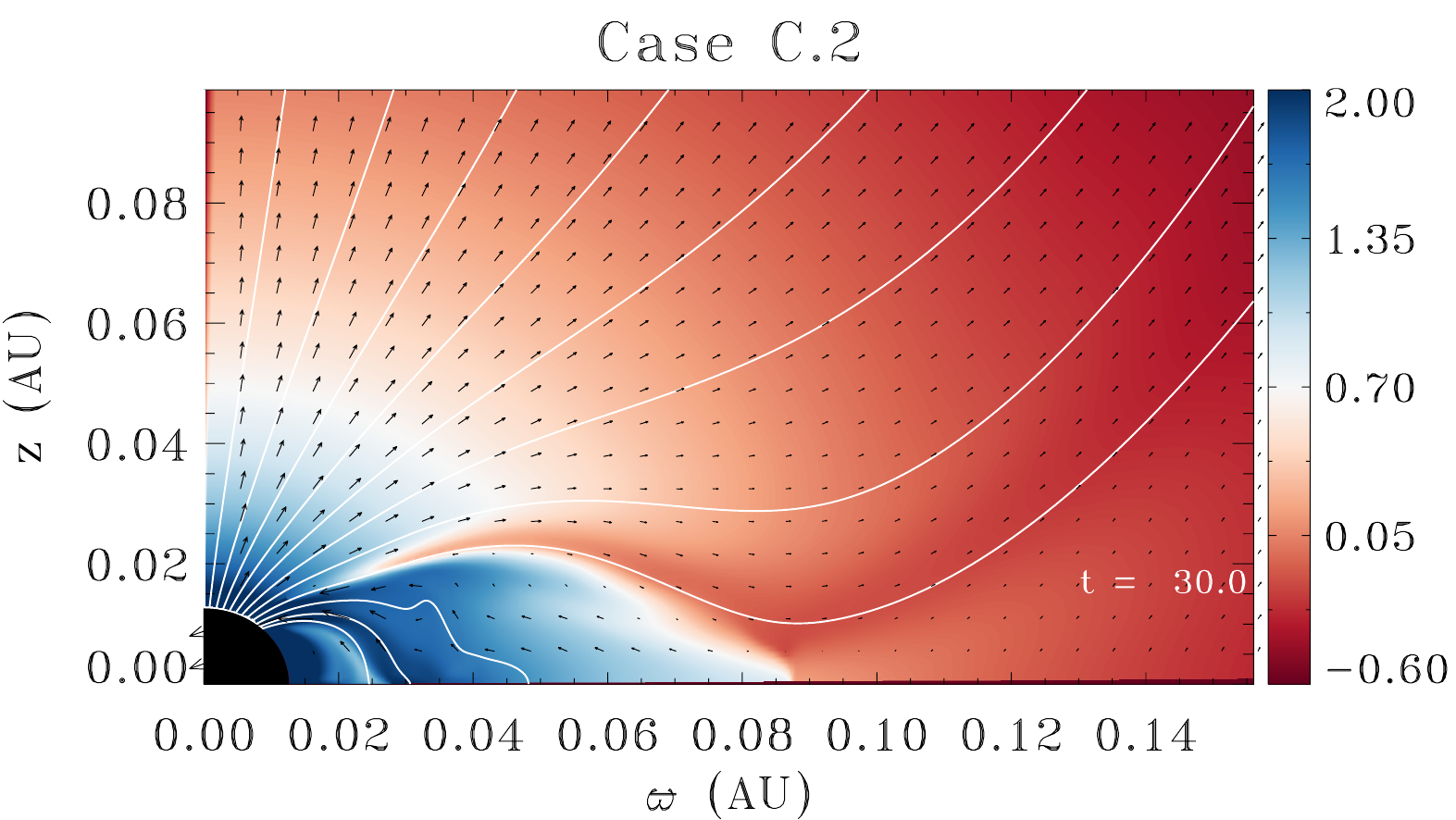} &   \includegraphics[width=0.48\linewidth]{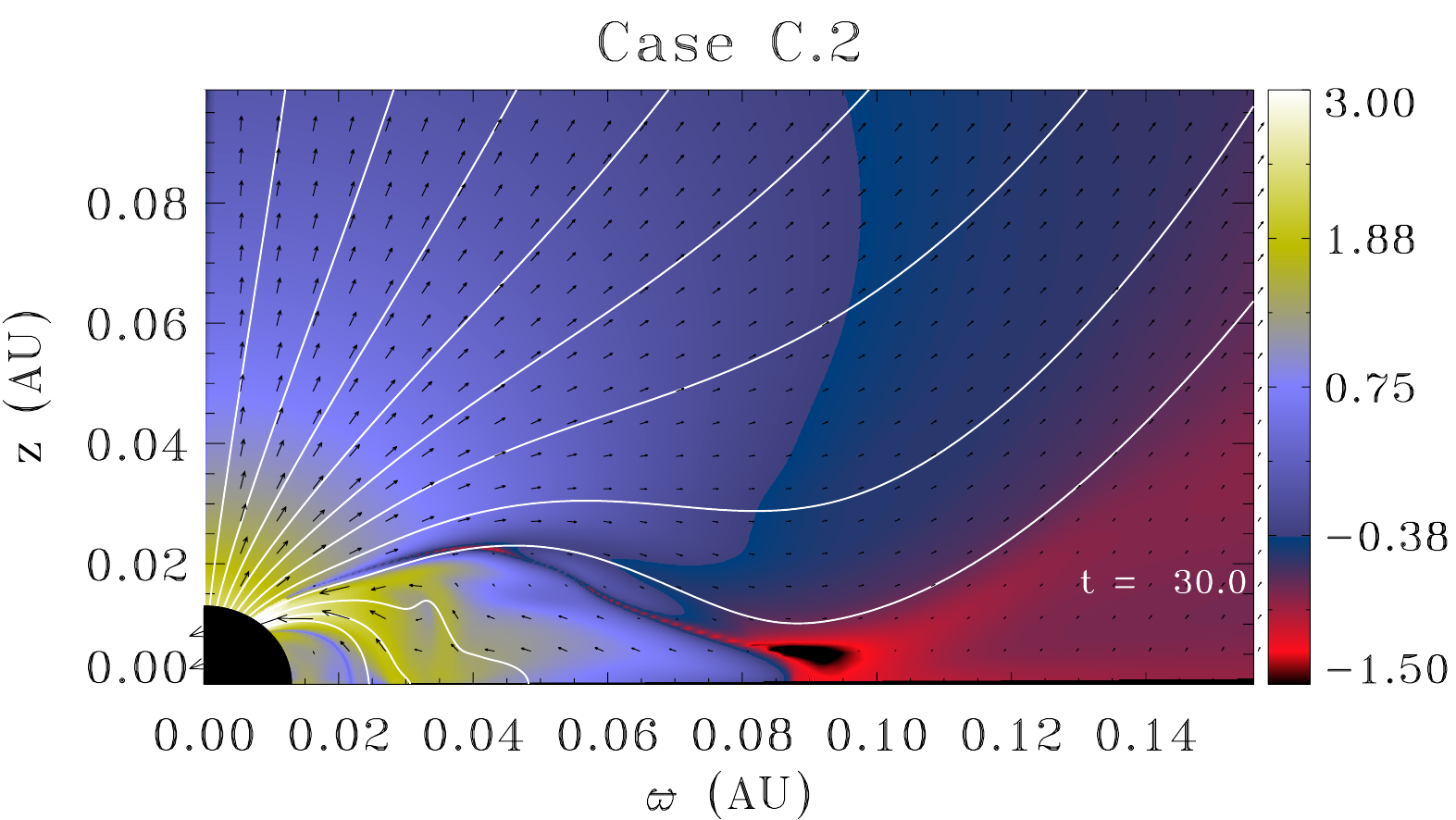} \\
(a) & (b) \\[6pt]
\end{tabular}
\caption{Plot of simulation case C.2, representative of the first solution type. Panel (a) shows the logarithmic density map in PLUTO units ($\rho_{\rm PLUTO}=2.48\times 10^{-15}$ g cm$^{-3}$) and panel (b) shows the logarithmic mass flux map, also in PLUTO units ($\rho V_{\rm PLUTO}=2.75\times 10^{-8}$  g cm$^{-2}$ s$^{-1}$).  The velocity vectors are shown as black arrows, and the magnetic field lines are represented by the solid white lines. The distances on the vertical ($z$) and horizontal ($\varpi$) axis are represented in astronomical units. The plot refers to PLUTO time = 30, which corresponds to 3.6 stellar rotations.} 
\label{FigC.2}
\end{figure*}
%--------------------------------------------------------

%We can see that inside the closed magnetospheric region the density is slightly lower for the first type of solutions, such as C.2 (Fig.\ref{FigC.2}), than for the second type, such as D.2 (Fig.\ref{FigD.2}) \textcolor{blue}{[But this was not expected? If we start with a large mass accretion rate! JFG]}. 

\begin{figure*}
\centering
\begin{tabular}{cc}
    \includegraphics[width=0.48\linewidth]{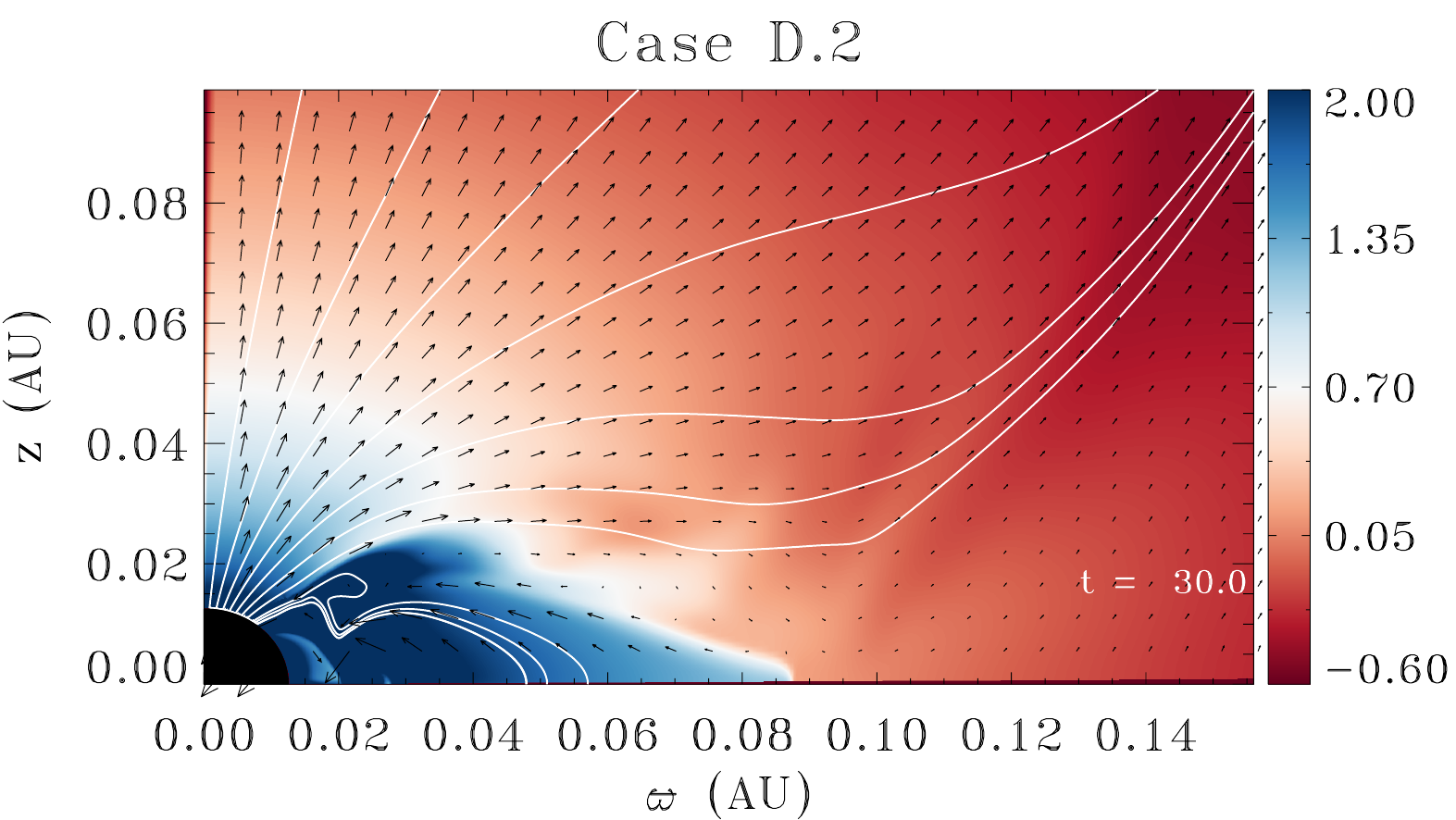} &   \includegraphics[width=0.48\linewidth]{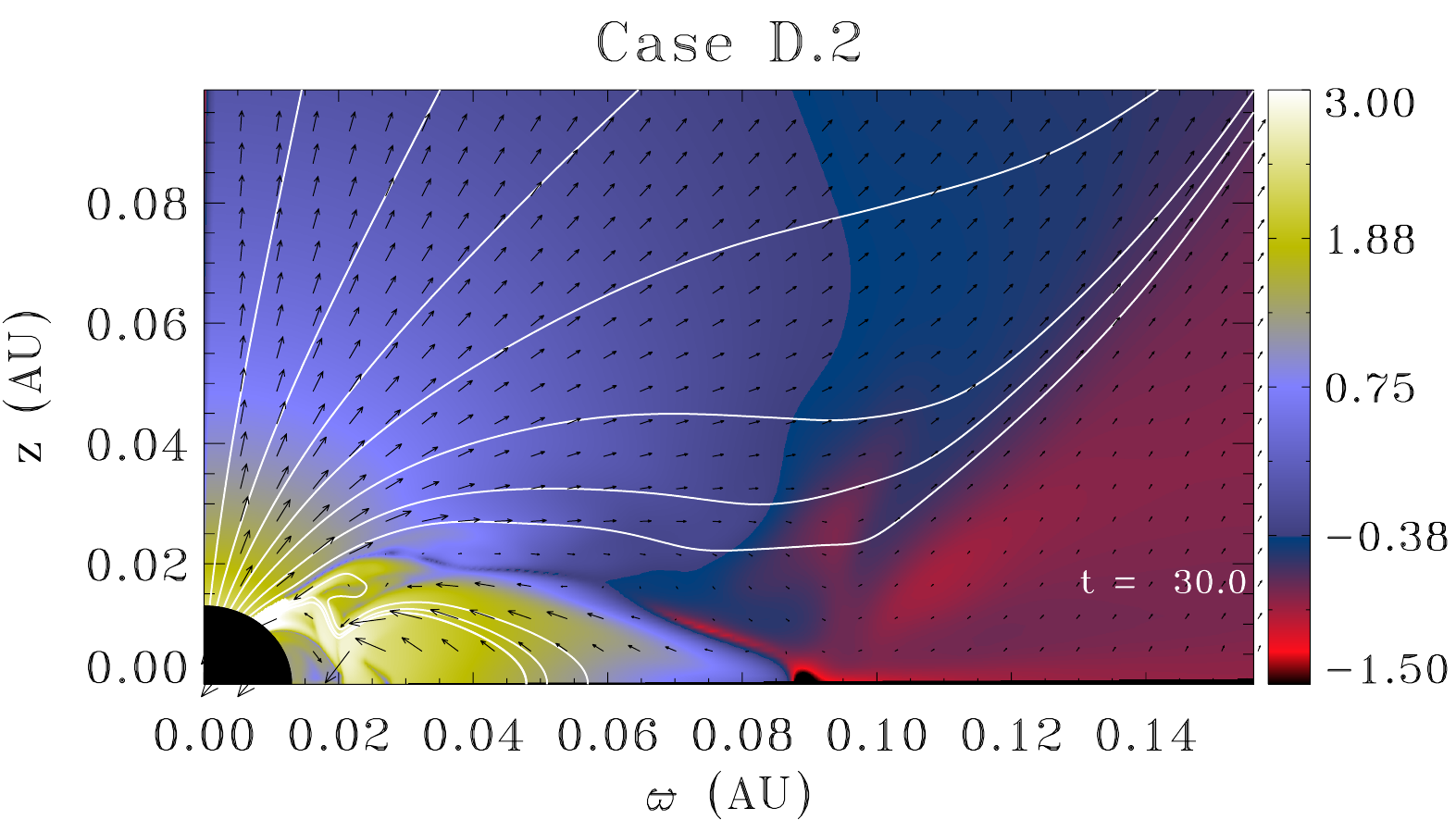} \\
(a) & (b) \\[6pt]
\end{tabular}
\caption{Plot of simulation case D.2, which is representative of the second solution type. Panel (a) shows the logarithmic density map in PLUTO units ($\rho_{\rm PLUTO}=2.48\times 10^{-15}$ g cm$^{-3}$) and panel (b) the logarithmic mass flux map, also in PLUTO units ($\rho V_{\rm PLUTO}=2.75\times 10^{-8}$  g cm$^{-2}$ s$^{-1}$).
 The velocity vectors are shown as black arrows, and the magnetic field lines are represented by the solid white lines. The distances on the vertical ($z$) and horizontal ($\varpi$) axis are represented in astronomical units. The plot refers to PLUTO time = 30, which corresponds to 3.6 stellar rotations.}
\label{FigD.2}
\end{figure*}

Simulation C.2 reaches a global fully steady configuration, in which a low-density channel separates the outflowing from the inflowing regions.  The stellar jet region in C.2 is even closer to the analytical solution than for D.2 when we compare Fig. \ref{FigC.2} (a) and Fig. \ref{FigD.2} (a) with the initial setup in Fig. \ref{Fig.t0}. The initial conditions are identical for solutions C.2 and D.2 in the outflowing region, and they have the same size.

Thus, by including a realistic inner stellar jet, we see that it is easily possible to completely suppress the magnetospheric ejection. Instead, a low-density region forms between the jet zone and the accretion zone. Because the density is so low in this intermediate zone, the magnetic field dominates. This region is therefore force free. Although this holds for jets from young stars, the result may have other applications to relativistic jets with two components, the inner spine jet and the outer disk wind. A configuration with an intermediate force free zone would help stabilize the inner leptonic jet from the black hole, embedded in the hadronic disk wind in the two stream models proposed by  \cite{Soletal89}.

Interestingly, the remarkable steadiness of the C-type solutions (C, C.1, and C.2) may be related to this low-density, static, force-free region. In this region, the magnetic pressure dominates the plasma pressure and leads to the formation of a low-density channel without evidence of outflowing material.

%--------------------------------------------------------
%           FIGURE D.2 FIG 4
%--------------------------------------------------------
%--------------------------------------------------------

%************************************************************
\subsection{Quasi-steady solutions with magnetospheric ejections} \label{sec:expected}
%************************************************************
 
Case D.2 is representative of the second type of simulations, as shown in Fig.\ref{FigD.2}. It shows a more complex but expected configuration. The inner stellar jet is very similar to the original analytical solution, but it is slightly more open with a lower mass-loss rate. Outside, the disk wind is close to the initial solution.  In addition, magnetospheric ejections are released by reconnection of the magnetic field lines, boosted by the increase in the accretion rate. This type of solution resembles the time-dependent MHD simulations performed by various authors, for example, \citet{ZanniFerreira13} and \citet{Romanovaetal09}. The approach used in our work emphasizes the non-negligible role of the inner stellar jet that stabilizes the magnetospheric ejection. 

Even though the mass seems to come from the star, the material of the  magnetospheric ejection eventually comes from the disk. The ejected mass from the disk bounces back onto the star, where the channel reduces the size.  This magnetospheric ejection would be observationally indistinguishable from an X-wind or an REX-Wind.  This release of material into the interstellar medium is accompanied  by consecutive disconnections and reconnections of the magnetic fields. Reconnection relies here on numerical diffusion. However, we verified that the timescale and the location of the reconnection events do not depend on the spatial resolution of the simulation. We are therefore confident about the physics of these events. We observed similar effects in previous simulations  (e.g., \citeauthor{Matsakosetal2012}, \citeyear{Matsakosetal2012}).

%************************************************************
\subsection{Varying the disk rotation profile} \label{sec:varyingDR}
%************************************************************

We performed several simulations with or without a Keplerian profile. In cases C.1, D.1, C.2, and D.2, we imposed Keplerian rotation on the equatorial plane from the last closed field-line outward (in the initial state, at $r=0.0875$ AU), as well as in disk wind region 4. Conversely, cases A, B, C, and D maintained the initial rotation profile of the analytical solution.  

By comparing case C, shown in Fig. \ref{FigTestABCD} (c), to case C.1, shown in Fig. \ref{FigtimeevolC.1} (c),  and case D, shown in Fig. \ref{FigTestABCD} (d), to case D.1, shown in Fig. \ref{FigtimeevolD.1} (g), we clearly see that the Keplerian disk wind profile does not affect the global structure. There might even be a slight stabilization effect of the stellar outflow because the Keplerian rotation profile produces an increase in the magnetocentrifugal launching. The disk wind density is slightly higher in this case, acting as a wall to maintain the stellar solution. 

We also ran similar simulations, imposing the Keplerian profile on the equatorial boundary alone, but keeping the analytical solution in the remaining box in the initial state. The overall solutions are again similar. 

Although we expected that the rotation discontinuity along the equatorial plane between the accretion disk and the magnetosphere would create strong instabilities, the code handled the sharp shear in rotational velocity quite well. The discontinuity smoothly diluted in the outflow.

%************************************************************
\subsection{Varying the size of the accretion zone and the dead zone} \label{sec:varyingAZ}
%************************************************************

%--------------------------------------------------------
%           FIGURE THIN ACCRETION ZONE
%--------------------------------------------------------
\begin{figure*}
\centering
\begin{tabular}{cc}
    \includegraphics[width=0.45\linewidth]{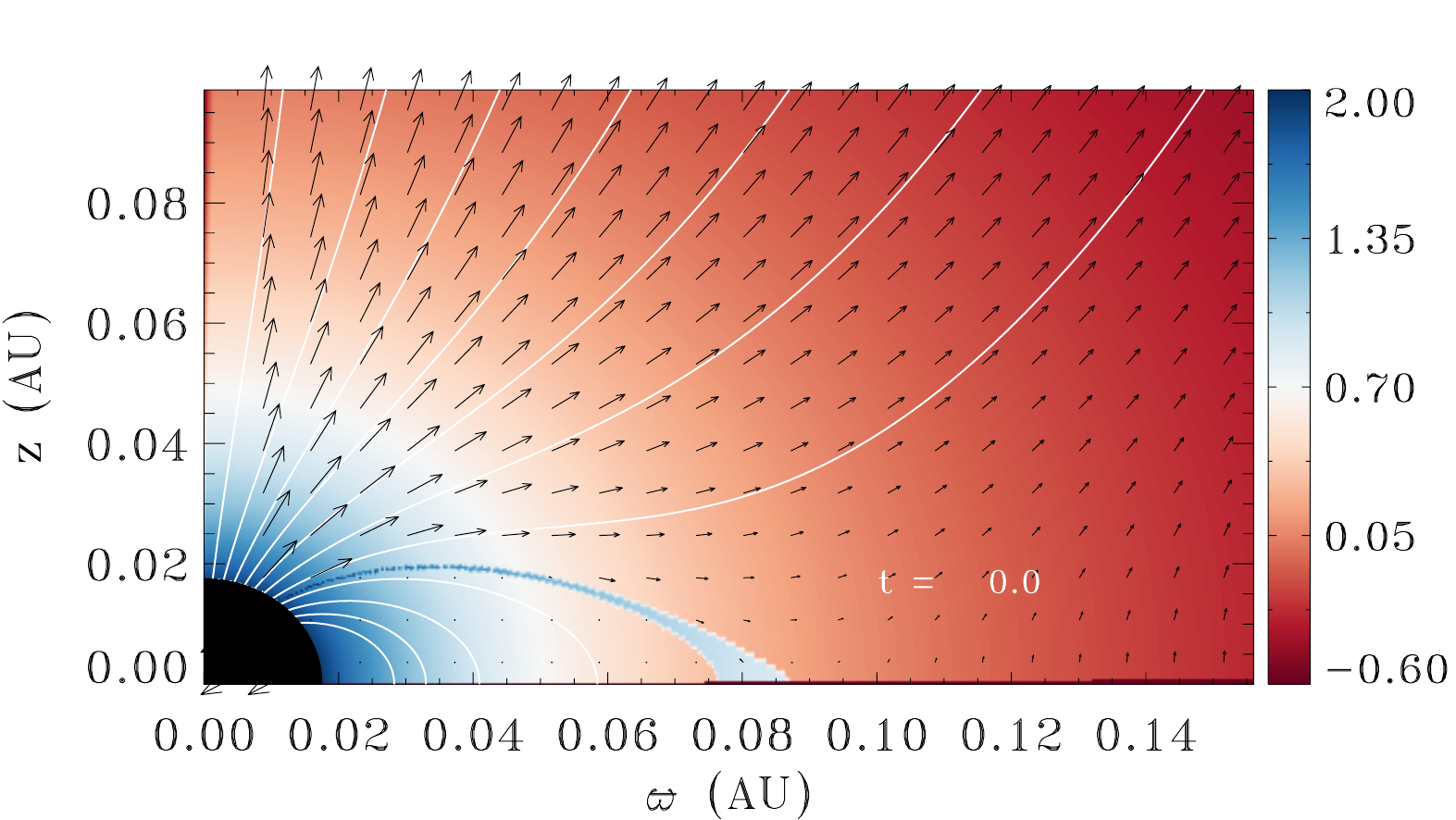} &   \includegraphics[width=0.45\linewidth]{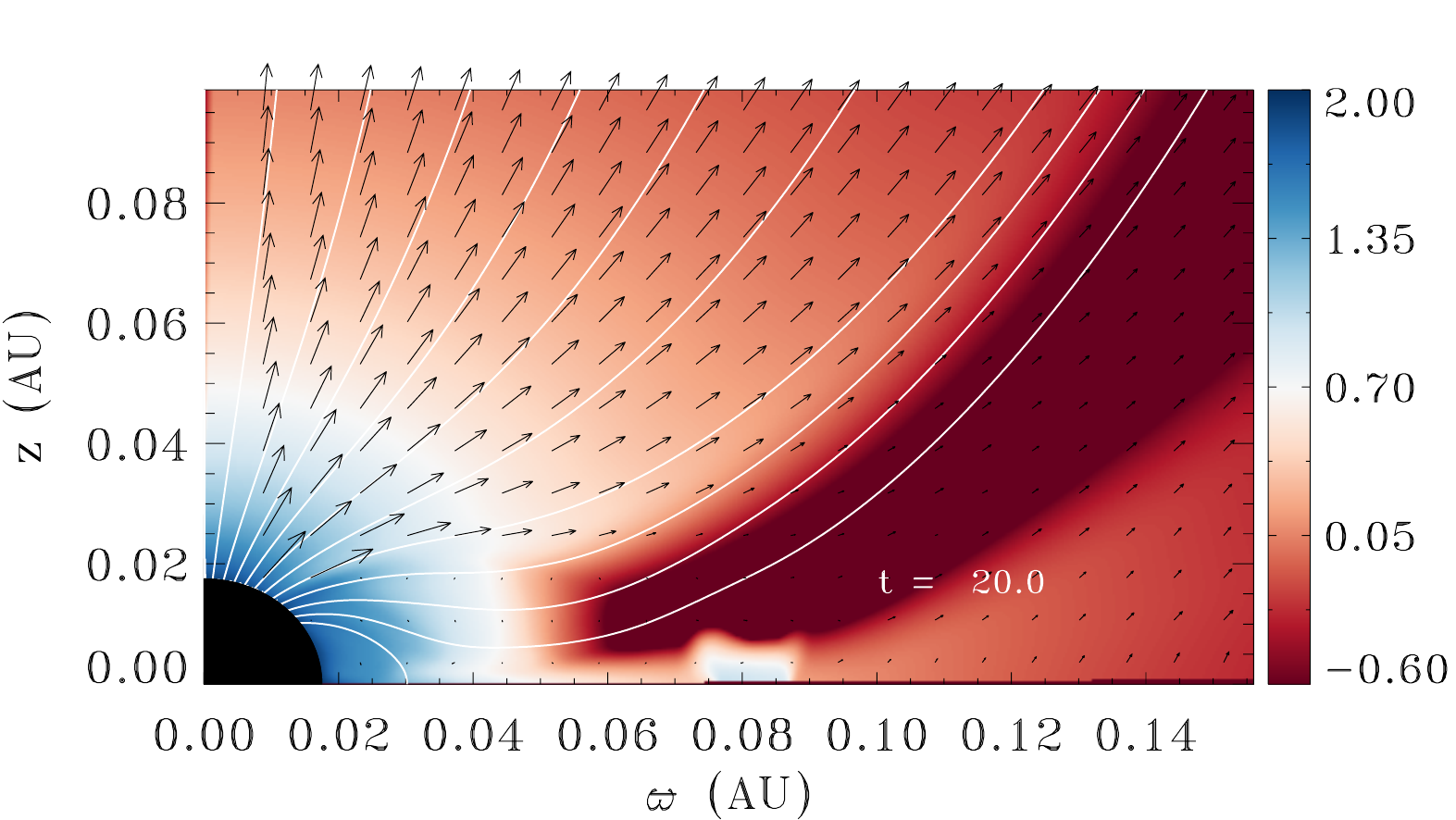} \\
(a) & (b) \\[6pt]
\end{tabular}
\caption{Density maps for a simulation with a very narrow initial accretion zone. Panel (a) shows the initial setup ($t=0$), and panel (b) corresponds to $t=20$ Pluto time units, i.e., 2.4 stellar rotations. The velocity vectors are shown as black arrows, and the magnetic field lines are represented by the solid white lines. The distances on the vertical ($z$) and horizontal ($\varpi$) axis are represented in astronomical units. 
\label{FigThin}}
\end{figure*}
%--------------------------------------------------------

We showed that the inner dead zone does not significantly affect the overall behavior of the ejected mass in the stellar jet or in the magnetospheric region.  The dead zone plays a quantitative role in controlling, by its size, the amount of accreted mass and the ejected magnetospheric mass-loss rate. It does not qualitatively change the fact that we still have two different types of simulations (C and D). For reasonable initial sizes of the dead zone, we observed that the system always stabilized with the same size of the final dead zone. This final size corresponds to the region where the equilibrium is force free.

We also performed other simulations with much larger dead zones, inducing a very thin accretion layer. This usually led to a disruption of the accretion column, and the entire disk material goes into the jet. In principle, it could correspond to a WTTS where accretion onto the star has supposedly stopped.
An example is given in Fig.\ref{FigThin}. It shows at $t=0$, Fig. \ref{FigThin} (a), a very thin accretion region where $\rho$ is multiplied by 5.0 and the velocity by $-1.5$, as in cases C, C.1, and C.2. At $t=20$ Pluto time units or 2.4 stellar rotations, Fig. \ref{FigThin} (b), an empty zone forms and the simulation is completely steady. The magnetospheric accretion has disappeared completely as well.

We do not assume that this is an exact simulation for a WTTS. It is more like a toy model, but it clearly indicates that we can perform stable and steady simulations where the stellar accretion stops, but the disk wind remains as well as the stellar jet. Observationally, such a configuration would have a very faint micro jet, but no large-scale powerful outflow. A full treatment of the accretion disk is needed in order to confirm this results. Here the disk is simply a boundary along the equatorial plane. 

%************************************************************
\subsection{Terminal jet and maximum accretion velocities} \label{sec:varyingAZ}
%************************************************************

%--------------------------------------------------------
%           FIGURE MF1 definition
%--------------------------------------------------------
\begin{figure}[tbp]
\includegraphics[width=\columnwidth]{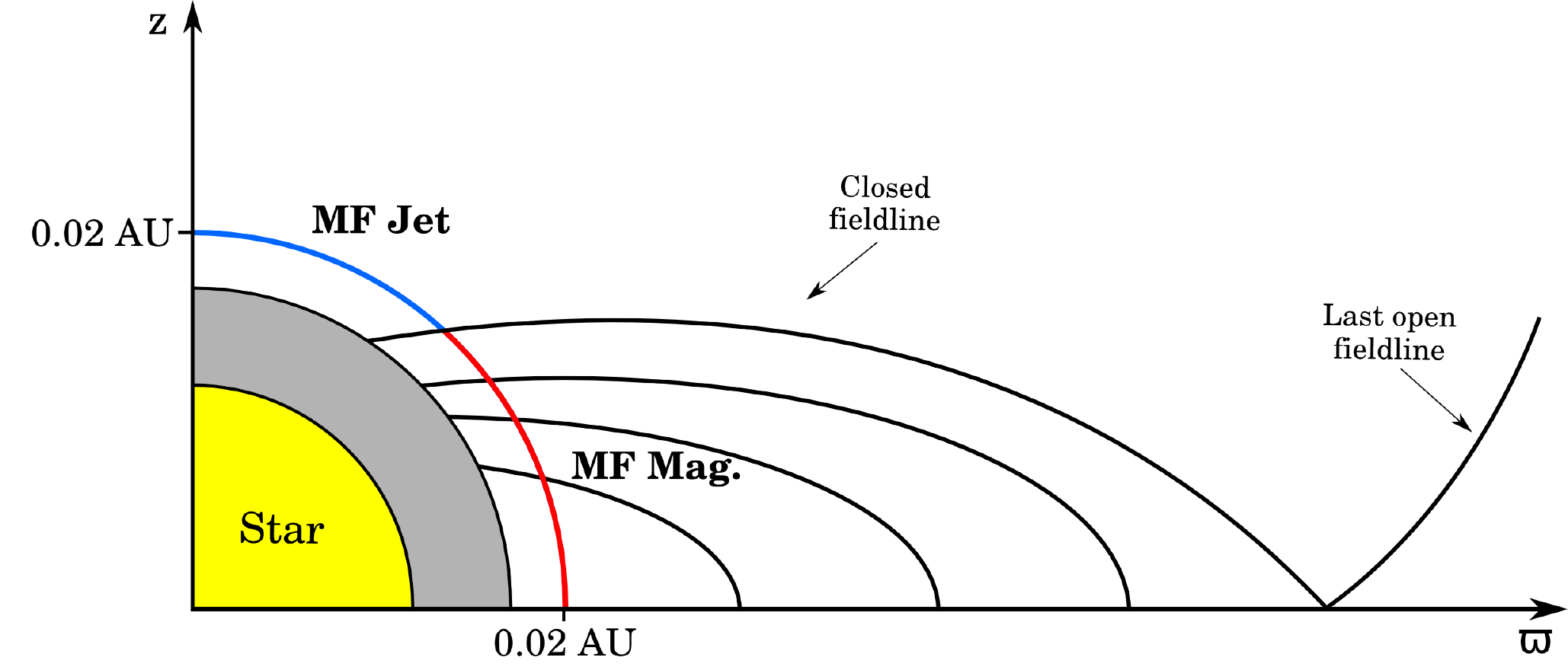}
\caption{Illustration of the regions in which the mass fluxes were measured. We calculate the jet mass-loss rate, MF Jet, along the blue line, where the velocity field is directed outward, and the magnetospheric mass-accretion rate, MF Mag, along the red line, where the velocity field is directed inward. The star is represented in yellow. The gray area extends from the surface of the star to $r_o$ , where integration starts. The magnetospheric region is delimited by the closed magnetic fieldlines (solid black lines), at the initial stage of the MHD simulations. 
\label{FigMF1def}}
\end{figure}
%---------------------------------------------------------------
%--------------------------------------------------------------------------------
% TABLE  2
%--------------------------------------------------------------------------------
\begin{table*}[htp]
\caption[Results for the simulations.]{Mass fluxes and velocities for the selected simulations. The simulations are identified in the first column. The next columns list the averaged mass-loss rates ($\dot M_{\rm loss}$) and mass-accretion rates ($\dot M_{\rm acc}$) in logarithmic scale, as well as the maximum accretion velocities ($V_{\rm acc}$) and the averaged terminal velocities of the jet projected along the line of sight ($V_{\rm proj}$).}
\label{tab:PLUTO_simulation_result}
%\resizebox{\textwidth}{!}{%
\centering
\begin{tabular}{lccccc}
\hline
Simulation ID & $\log \dot{M}_{\rm loss}$ & $\log \dot{M}_{\rm acc}$ & $\dot{M}_{\rm loss} /\dot{M}_{\rm acc}$ & $V_{\rm acc}$ & $V_{\rm proj}$ \\
 & ($\Msunyr$) & ($\Msunyr$) &  & (\kms) & (\kms) \\
\hline
Case A & $-8.6$ & $-8.4$ & $0.71$ & $95$ & $138$ \\
Case B & $-8.6$ & $-8.3$ & $0.45$ & $151$ & $138$ \\
Case C & $-8.6$ & $-7.7$ & $0.12$ & $312$ & $137$ \\
Case C.1 & $-8.6$ & $-7.9$ & $0.22$ & $289$ & $143$ \\
Case C.2 & $-8.6$ & $-7.7$ & $0.13$ & $295$ & $163$ \\
Case D & $-8.6$ & $-7.3$ & $0.05$ & $225$ & $138$ \\
Case D.1 & $-8.7$ & $-7.8$ & $0.11$ & $226$ & $143$ \\
Case D.2 & $-8.6$ & $-7.3$ & $0.05$ & $215$ & $158$ \\
\hline
\end{tabular}
%}
\end{table*}
%--------------------------------------------------------------------------------

To characterize the inflow and outflow dynamics, we calculated the maximum accretion velocity inside the closed magnetosphere and the terminal velocity of the stellar jet projected along the line of sight.  The results are summarized in the two last columns in Tab. \ref{tab:PLUTO_simulation_result}.

The terminal velocities of the stellar jet were calculated by averaging along a vertical line between $z=2.2$ and $z=2.75$ AU and at the cylindrical radius of $\varpi\approx 0.8$ AU, which corresponds to the radius of the stellar jet.
Then, taking for RY Tau an inclination  of $i=65^o$ (\citeauthor{Long2019}, \citeyear{Long2019}), we obtained  values between $137$ \kms and $163$ \kms\  for the projected velocities along the
line of sight. This is consistent with observations. All these values are so close to each other that it would be impossible to observationally separate the different cases because velocities are usually measured with an uncertainty between 10 and 20\%.

The projected velocities of cases C.2 and D.2, $163$ \kms and $158$ \kms , correspond to an averaged terminal velocity of the jet of $385$ \kms and $374$ \kms, respectively. These are the simulations with the highest resolution. These velocities are close to the value of the initial analytical solution, which has a terminal velocity along the polar axis of $393$ \kms\  (see \citeauthor{sautyetal11}, \citeyear{sautyetal11}).

Simulations C, C.1, and C.2 show higher maximum accretion velocities than simulations D, D.1, and D.2, see Tab. \ref{tab:PLUTO_simulation_result}. For instance, in case C.2, we measure a maximum accretion velocity of $295$ \kms , while in case D.2,  the maximum accretion velocity is $215$ \kms. This result may sound counterintuitive. One possible explanation is the following. In case C.2, the mass-accretion flux is higher between 0.02 and 0.05 AU on the equatorial plane, as shown in Fig. \ref{FigC.2} (b) in green. In case D.2, the mass accretion flux is spread over a wider range of radii between 0.02 and 0.07 AU on the equatorial plane, shown in Fig. \ref{FigD.2} (b) in green. As a consequence, in case D.2, the mass with higher velocities coming from the disk between 0.06 and 0.08 AU does not reach the star, but feeds the magnetospheric ejection, which may explain the lower maximum accretion velocity we finally obtain.

%************************************************************
\subsection{Mass accretion and mass-loss rates} \label{sec:massflux}
%************************************************************
%--------------------------------------------------------
%           FIGURE Mass FLUX A,B,C,D,C.1,D.1
%--------------------------------------------------------
\begin{figure*}
\includegraphics[width=1.0\hsize]{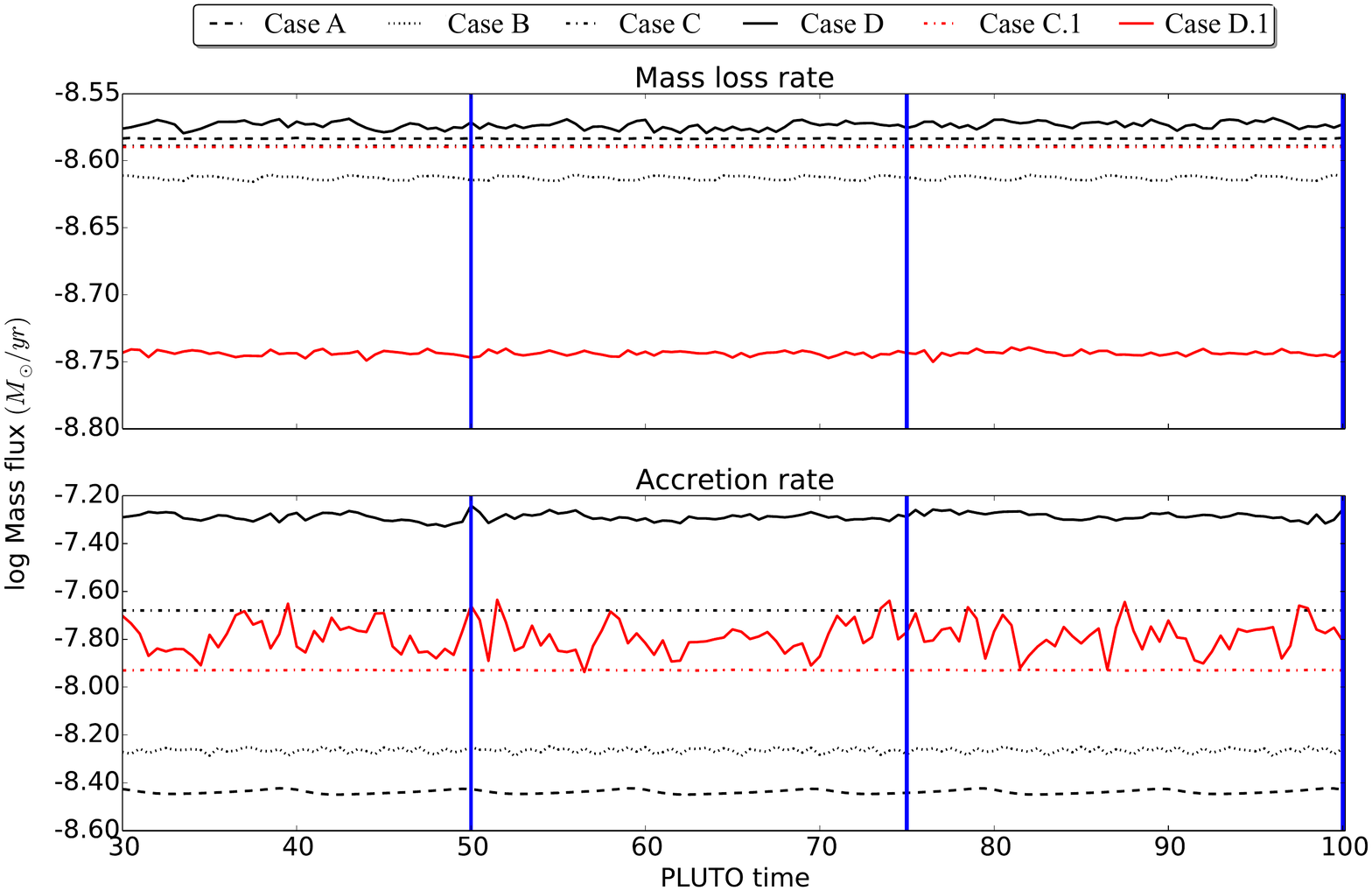}
\caption{Logarithmic mass flux values measured in the stellar jet and accreting magnetospheric regions for cases A, B, C, D, C.1, and D.1 between PLUTO time 30 and 100, which means between $3.6$ and $12$ stellar rotations. The vertical blue lines correspond to PLUTO times 50, 75, and 100 used for the plots of case D.1 in Fig. \ref{FigtimeevolD.1}. }
\label{FigMF1ABCD}
\end{figure*}
%--------------------------------------------------------

The mass fluxes were calculated for each PLUTO time in two regions onto a spherical shell of  radius $0.02$ AU. We calculated the mass-loss rate, $\dot{M}_{loss}$,  in the stellar jet region (region 1 of Fig. \ref{Fig.t0}) using the spherical shell MF Jet in Fig. \ref{FigMF1def} and the mass-accretion rate, $\dot{M}_{acc}$, in the closed magnetospheric region (region 2 of Fig. \ref{Fig.t0}) using the spherical shell MF Mag in Fig. \ref{FigMF1def}. 

To derive the mass fluxes, we integrated $\rho V_p$ across the open magnetic field lines of the stellar jet region for $\dot{M}_{loss}$ and across the foot points of the closed magnetospheric flux tubes for $\dot{M}_{acc}$. Tab. \ref{tab:PLUTO_simulation_result} gives the averaged mass-accretion rate, the mass-loss rate, the ratio $\dot{M}_{\rm loss} /\dot{M}_{\rm acc}$, the maximum of the accretion velocity, and the averaged terminal velocity of the jet  for the eight selected simulations.

\begin{figure}
\includegraphics[width=
\hsize]{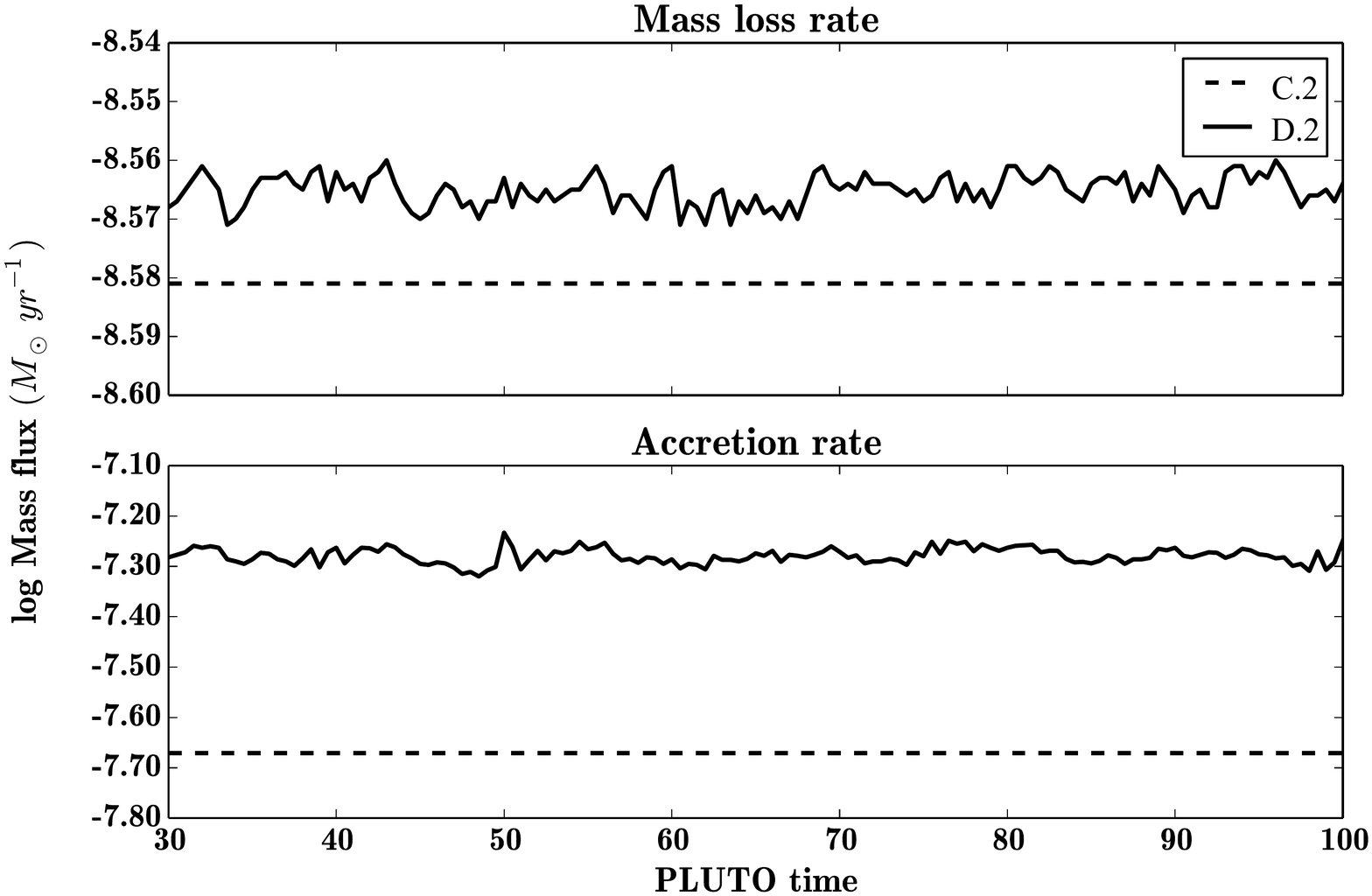}
\caption{
Logarithmic mass flux values measured in the stellar jet and accreting magnetospheric regions for cases C.2 (dashed line) and D.2 (solid line) between PLUTO time 30 and 100
($3.6$ and $12$ stellar rotations, respectively).
 }
\label{FigMF1C.2D.2}
\end{figure}

We plot the measured mass fluxes with time for cases A, B, C, D, C.1, and D.1 in Fig.\ref{FigMF1ABCD} and  for the two cases C.2 and D.2 in Fig.\ref{FigMF1C.2D.2}. All plots extend from PLUTO time 30 to 100.  Basically, all simulations have very similar constant stellar mass-loss rates. This figure confirms the steadiness of cases C, C.1 and C.2. Cases D, D.1, and D.2 have a slightly higher variation in the mass-loss rates. When Case D.2 is taken as an example, the stellar mass-loss rate, $\dot{M}_{loss}$, varies from $2.69{\,\times}\,10^{-9}$  to $2.75{ \,\times}\,10^{-9} \Msunyr$ , however, which is a variation of less than 2\%, which is negligible. 
%(e.g. $10^{-8.58}=2.63{ \,\times}\,10^{-9} \Msunyr$ in Case C.2 and $10^{-8.56}=2.74{ \,\times}\,10^{-9} \Msunyr$ in Case C.1) 
The averaged mass-loss rate of all our simulations is very close to the value of the analytical solution, $10^{-8.5}=3.16{ \,\times}\,10^{-9} \Msunyr$, taken from \cite{gomezetal01}. 

 Cases A and B are not relevant for the mass-accretion rates because they are so low that the ratio of mass-loss rate to mass-accretion rate is unrealistic. Mass-accretion rates are constant in cases C, C.1, and C.2. The value for C.1 is slightly lower than the value for C because the dead zone suppresses part of the accretion onto the star. The difference in accretion rate between case C.1 and case C.2 is related to the higher resolution of the second case. The averaged values of the mass-accretion rates of cases D, D.1, and D.2 follow the same trend.
The time-averaged mass-accretion rates are $10^{-7.68}=2.1{ \,\times}\,10^{-8}\Msunyr$ for C.2 and $10^{-7.27} =5.3{\,\times}\,10^{-8}\Msunyr$ for D.2,  in agreement with the typical range of values found for RY Tau  \citep{ hartiganetal95, calvetetal04, mendigutiaetal11, costiganetal14}. In any case, the differences in the averaged mass-accretion rates of simulations C, C.1, C.2, D, D.1, and D.2 are well below the observational precision.

We note that cases D, D.1 and D.2 show stronger variation of the mass accretion rate than cases A, B, C, C.1, and C.2 because of the magnetospheric ejection. This is precisely the variation in the mass-accretion rate onto the star that can help us to estimate the mass loss in the magnetospheric ejection.  In case D, the mass-accretion rate, $\dot{M}_{acc}$, varies from $4.4{ \,\times}\,10^{-8}$  to $5.6{ \,\times}\,10^{-8} \Msunyr$, which means a variation of 27\%. In case D.1, it varies from $1.26{ \,\times}\,10^{-8}$  to $2.5{ \,\times}\,10^{-8} \Msunyr$, which represents a variation of 98\%. In case D.2, the mass-accretion rate onto the star, $\dot{M}_{acc}$, varies from $5.01{ \,\times}\,10^{-8}$  to $6.30{ \,\times}\,10^{-8} \Msunyr$, which means a variation of 26\%.

Because we fixed the boundaries on the equatorial plane, the equatorial mass-accretion flux from the disk remains constant. Because the mass has to go somewhere, the difference between the minimum and the maximum of the mass accretion rate onto the star corresponds to the mass-loss rate of the magnetospheric ejection. The magnetospheric mass-loss rate of case D peaks at $1.19{ \,\times}\,10^{-8} \Msunyr$ , while the average stellar mass-loss rate is around $2.7{ \,\times}\,10^{-9} \Msunyr$. For case D.1, the equivalent magnetospheric mass-loss rate peaks at $1.23{ \,\times}\,10^{-8} \Msunyr$ , while the average stellar mass-loss rate is about $1.8{ \,\times}\,10^{-9} \Msunyr$.
For case D.2, this magnetospheric mass-loss rate peaks at about $1.30 {\,\times}\,10^{-8} \Msunyr$, which is again roughly four times the stellar mass-loss rate. However, in all cases, the magnetospheric mass-loss rate is strongly variable. In total, the episodic magnetospheric ejection mass is therefore comparable in magnitude to the continuous stellar mass-loss rate. This means that the variation may not be observationally detectable, considering that we measure the total mass-loss rate, the stellar component, and the magnetospheric component, and mass-loss rates are usually determined within a factor of about 2.

%--------------------------------------------------------
%           FIGURE Mass FLUX C.2 D.2
%--------------------------------------------------------
%--------------------------------------------------------

Although the difference between the total mass-loss rate does not vary by more than a factor of 2 on average, the ejected mass from the magnetosphere in cases D, D.1, and D.2 is high enough to  significantly clear the environment of the star and decrease the circumstellar extinction. This fits the scenario proposed by \cite{petrovetal19} to explain the two states of RY Tau. Case D.2 would correspond to the active state with  an enhanced outflow, as we explain in Sec. \ref{sec:observations}.

%*******************************************************************
\section{Discussion} \label{sec:discussion}
%********************************************************
%********************************************************
\subsection{Comparison with other simulations} \label{sec:Comparison}
%********************************************************

We performed simulations of the close environment of RY Tau with initial conditions corresponding to the analytical solution of \cite{sautyetal11}.
We started by including a magnetospheric accretion region alone, increasing the mass-accretion rate. Then, the simulation set up was improved by the inclusion of a dead zone and a Keplerian rotation profile in the disk. Finally, we increased the numerical resolution.

We reached a quasi-steady state within 3 stellar rotations, which corresponds to 42.6 days for the star of our simulations. \cite{ZanniFerreira13} reached stable equilibria after more than 50 stellar rotations, which corresponds to more than 250 days, because their star rotated faster. \cite{Romanovaetal09, Romanovaetal11} reached stable equilibria after 500 days, which corresponds to about 100 stellar rotations in their case. Thus, our method reduces the relaxation timescale by a factor of 5 to 10. More recent simulations of \cite{Irelandetal2021} achieved stable equilibria after 2 to 3 rotations of the central star. The reason probably is that they also included a stellar jet component with a high velocity, but very low density. However, it should be noted that all other simulations use smaller grids with lower resolutions (see the end of Sec. \ref{subsec:numericalsetup} for a detailed comparison). Another reason for our shorter simulation time may be that we started close to equilibrium because we used the analytical solution as initial conditions.

By increasing the mass-accretion rate in our simulations, we found two classes of solutions with different behaviors. Class 1 corresponds to lower mass-accretion rates and class 2 to higher mass-accretion rates. For class 2, we have quasi-steady solutions with magnetospheric ejections (Cases D, D.1, and D.2). They are somewhat similar to the simulations presented by \cite{ZanniFerreira13} and \cite{Romanovaetal09} with a conical magnetospheric mass ejection, which removes some of the angular momentum from the star and the disk. Class 1 is a new type of steady simulations (cases A, B, C, C.1, and C.2) that are different from those of \cite{ZanniFerreira13}, \cite{Ustyugovaetal2006} and \cite{Romanovaetal09}, but closer to the stellar jet simulations of \cite{MattPudritz05,MattPudritz08}. Class 1 simulations do not show episodic magnetospheric ejecta, but instead a low-density channel separating the magnetospheric accretion zone from the stellar jet. This may be due to the presence of a strong stellar component.
The low-density force-free region is more pronounced for higher accretion rates such as case C than cases A or B. However, the accretion rates in cases A and B are very close to the stellar mass-loss rates (see Tab. \ref{tab:PLUTO_simulation_result}). However, these ratios close to one are not observed in YSOs, and we therefore did not apply these two cases to RY Tau.

Clearly, the transition from class 1 to class 2 is 
between case C and case D, without a dead zone (see Fig. \ref{FigTestABCD}), between case C.1 (Fig. \ref{FigtimeevolC.1}) and case D.1 (Fig. \ref{FigtimeevolD.1}) with a dead zone, and between case C.2 (Fig. \ref{FigtimeevolC.1}) and case D.2 (Fig. \ref{FigtimeevolD.1}) with a dead zone and an increased accuracy. There is a similar transition between these two classes, regardless of the size of the dead zone and of the inner boundary of the simulation. Tab. \ref{tab:PLUTO_simulation_result} shows that class 2 solutions only arise when the ratio of mass loss to mass-accretion rate is lower than 0.11. 

These two classes of solutions we found have similar mass-loss rates. 
When we compared the two classes of solutions, we obtained a stronger variation in mass-accretion rate than of mass loss (e.g., cases C.2 and D.2 presented in Fig. \ref{FigMF1C.2D.2}), but from an observational point of view, it would be very difficult two distinguish the two classes because the uncertainties on the measurements of mass-loss rates and mass-accretion rates are much higher than the differences found in our simulations. 

Our two types of solutions can be compared to the propeller regime versus conical wind regime observed by \cite{Ustyugovaetal2006} and \cite{Romanovaetal09}  in their simulations. As mentioned before, it is slightly different from our study. \cite{Ustyugovaetal2006} and \cite{Romanovaetal09} always had magnetospheric ejection from the disk in the form of a massive conical wind. Only in the propeller regime is there a strong axial jet from the star because the rotation is strong and much closer to the breakup velocity. This simulation therefore applies to fast-rotating stars. As a consequence, the propeller and the conical wind regimes cannot be observed simultaneously in a star, as the stellar rotation evolves over thousands of years.

\cite{ZanniFerreira13} claimed in their conclusion that the mass flux in their magnetospheric ejections is 1\%-2\%\  of the mass-accretion rate. As we mentioned, the mass-loss rate in the magnetospheric mass flux of D.2 varies up to 26 \% of the mass-accretion rate, while the stellar mass-loss rate is of about 5\%. 
Thus the ratio of the mass ejected from the magnetosphere to the accreted mass on the start is comparable to the ratio obtained in \cite{ZanniFerreira13} and also in
\cite{Romanovaetal09} 
A more detailed analysis of the magnetospheric ejecta can be found in \cite{Romanovaetal11}
and \cite{Irelandetal2021}, but the qualitative conclusions are identical for the comparison with our simulations.

Our simulations do not differ from the analytical solution for the stellar spin-down. The mass-loss rate and the angular momentum flux extracted by the stellar jet are similar in the analytical solution and numerical simulations.  Using the method explained \cite{sautyetal11}, we calculated a braking time of about 0.6 million year or even shorter. There are two main reasons for this similarity between the analytical solution and the numerical simulations. First, in \cite{sautyetal11}, only the stellar jet open field-line region  was responsible for the stellar braking. Second, the inversion of the toroidal magnetic field, $B_\varphi$, and the poloidal velocity, $V_{\rm P}$, in the magnetospheric accretion zone ensures that the total angular momentum flux
on the star remains positive, directed outward. Thus the disk effectively magnetically brakes the star.
In other words, the magnetospheric field-lines exert an efficient disk-locking on the star. The end result is that the stellar jet and magnetospheric accretion spin-down the star in a similar way in our simulations. As in \cite{sautyetal11}, this gives a spin-down of less than one million years for the star, which is well within the lifetime in the CTTS phase.

%********************************************************
\subsection{Comparison with observations of RY Tau} \label{sec:observations}
%********************************************************

The literature contains many values for the mass loss and accretion rates for RY\ Tau determined with different methods throughout the years. The values for the mass-accretion rates vary from $10^{-7.7}$ to $10^{-7.0} \Msunyr$, while the mass-loss rate oscillates between $10^{-8.8}$ and $10^{-7.1} \Msunyr$ \citep{kuhi64,edwardsetal87,hartiganetal95,gomezetal01,agra-amboageetal09,Skinneretal18}.

In the following, we concentrate our discussion on cases C.2 and D.2 as they correspond to the most extended grids and the highest resolution.
The mass-loss rates for these two cases do not deviate significantly from the value given by \citet{gomezetal01}, $10^{-8.5} \Msunyr$, for the RY Tau microjet.  Case C.2 agrees with the value of $10^{-7.7} \Msunyr$ obtained by \citet{mendigutiaetal11} for the mass-accretion rates, and  the results of case D.2 are within the range of values derived by \citet{costiganetal14}, namely $10^{-7.6}-10^{-7.1} \Msunyr$. The mass fluxes obtained in these two classes of solutions are well within the measurement error bars in any case, which means that it difficult to determine which case fits the observations best.

Further constraints could be achieved by comparing the ratio of the mass loss and accretion rates. For RY Tau, \citet{agra-amboageetal09} estimated values ranging from 0.02 and 0.4 and \citet{hartiganetal95} reported a ratio  of $0.06$.  When compared with our simulations, which show ratios of $0.13$ and $0.05$ for cases C.2 and D.2, respectively, the quasi-steady solution of case D.2 is consistent with the latter reference. The broad range of values observed by \citet{agra-amboageetal09} do not allow us to distinguish between the two cases C.2 and D.2.

The maximum accretion velocities measured in the simulations are within the range of velocities mentioned by \citet{edwardsetal87}, between 200 and 300 \kms.
As mentioned, we retrieved projected velocities in the line of sight of the observer of $162$ \kms for case C.2, and $158$ \kms for case D.2. Both values are compatible with the velocity  of $-136 \pm 10$ \kms \ obtained by  \citet{Skinneretal18} from the observation of the C IV line at a distance of 39 AU from the star.

In a recent analysis of simultaneous spectral and photometric observations carried out for 5 years, \cite{petrovetal19}  found that  RY Tau has a bimodal behavior. For most of the time, the star exhibited a variable outflow, although a period of quiescent state was also observed. This latest state was observed for several months, in which this star became fainter and less active in outflowing activity.  Conversely, periods of higher brightness (lower circumstellar extinction) of RY Tau are correlated with moments of increased outflow activity observed from the spectral profile of H$\alpha$. \cite{petrovetal19} suggested that these outflows might be sporadic magnetospheric ejections that modify the dusty disk wind geometry and therefore the brightness of the star and spectral footprint. We propose that the steady configuration of case C.2 may match the quiescent epoch of RY Tau  and the configuration of case D.2 may fit the active stage. We suggest that the increase in the outflow intensity during the episodic plasma ejections is a consequence of an increase in accretion, as shown in the simulations from the transition of cases C.2 to D.2. \cite{petrovetal19} found that although the instant accretion rate is highly variable, the average level does not change significantly between the quiescent and the active regime. Unfortunately, the mass-accretion rate transition between C.2 and D.2 in our simulations falls within the uncertainty on the average observed mass-accretion rate. We therefore cannot reach a definitive conclusion for the case of RY Tau.

%***********************************************************************
\section{Conclusion \label{sec:conclusion}}
%***********************************************************************

We used a pressure-driven outflow model with accretion implemented in the closed magnetosphere to simulate the accretion and outflowing environment of the intermediate-mass classical T Tauri star RY Tau. Clearly, our best simulations correspond to cases C.2 and D.2.
The increase in velocity and density from cases C.2 to D.2 enables us to obtain higher accretion rates.  Furthermore, when we computed the ratios of ejected and accreted material, the values of the two simulations are quite consistent with the literature and support the observational evidence that accretion dominates ejection.

Additionally, we were able to achieve a steady configuration for class 1 solutions in 2.5 stellar rotations, in contrast to all previous simulations that did not take the coronal stellar wind properly into account. The more recent simulations of \cite{Irelandetal2021} have a stellar jet component and comparable timescales. 

Class 2 simulations present a more perturbed configuration in which magnetospheric ejections propagate between the stellar jet and closed magnetospheric regions. Although comparable, the magnetospheric mass-loss rate does not dominate the stellar mass-loss rate. Much higher accretion rates are needed for a strong magnetospheric flow.
The accretion velocities measured for cases C.2 and D.2 seem to agree with observations, as do the terminal velocities determined for the stellar jet. 

We suggest that the classes of solutions obtained in this study match the two behaviors observed for RY Tau in \cite{petrovetal19}. 
C.2 could match the quiescent epoch of RY Tau, characterized by lower brightness, and D.2 could be linked to a more active stage that is characterized by enhanced outflow activity. 

In a future work, the analytical solution used in the simulations could be extended to another range of stellar radii and masses in order to replicate the circumstellar environment of other CTTS. We also started new simulations initiating the disk wind solution with an analytical Blandford and Payne-type solution \citep{BlandfordPayne1982}. This is quite promising, but the present work is an essential preliminary study before we proceed with more sophisticated models.

%*********************************************************************************
\begin{acknowledgements}
%*********************************************************************************

This work was supported by Funda\c{c}\~ao para a Ci\^encia e a Tecnologia (FCT) through national funds and by Fundo Europeu de Desenvolvimento Regional (FEDER) through COMPETE2020 - Programa Operacional Competitividade e Internacionaliza\c{c}\~ao by these grants: UIDB/04434/2020; UIDP/04434/2020;  UID/FIS/04434/2019; PTDC/FIS-AST/32113/2017 \& POCI-01-0145-FEDER-032113.
RMGA acknowledges support from FCT through the Fellowship PD/BD/113745/2015 (PhD::SPACE Doctoral Network PD/00040/2012) and POCH/FSE (EC).
CS thanks LUPM for hosting him during his CNRS delegation and his "CRCT". 
We acknowledge financial support from ``Programme National de Physique Stellaire'' (PNPS) and from "Programme National des Hautes Energies" (PNHE) of CNRS/INSU, France, and from CRUP through a cooperation program (PAULIF: TC-16/17). We are grateful to A. Mignone and collaborators for letting us use the PLUTO code (http://plutocode.ph.unito.it, version 4.2). In particular we thank T. Matsakos for his useful and valuable recommendations in using the code. 
We also thank P.P. Petrov, C. Meskini, N. Vlahakis, K. Tsinganos and an anonymous referee for valuable comments.
\end{acknowledgements}

%*********************************************************************************
% REFERENCES
%*********************************************************************************

\bibliographystyle{aa}
\bibliography{references}

\end{document}